\documentclass[conference, anonymous]{IEEEtran}
\renewcommand\IEEEkeywordsname{Index terms}
\usepackage{color}
\usepackage[english]{babel}
\usepackage[utf8]{inputenc}
\usepackage{algorithm}
\usepackage{subcaption}
\usepackage{adjustbox}
\usepackage[noend]{algpseudocode}
\usepackage{float}
\usepackage[noend]{algpseudocode}
\usepackage{graphicx}
\usepackage{scalerel}
\usepackage{wasysym}
\usepackage{tabu}
\usepackage{marvosym}
\usepackage{caption}
\usepackage{subcaption}
\usepackage{xcolor}
\usepackage{lipsum}
\usepackage[noadjust]{cite}
\usepackage{tikz}
\usepackage{calc}
\def\checkmark{\tikz\fill[scale=0.4](0,.35) -- (.25,0) -- (1,.7) -- (.25,.15) -- cycle;} 

\usepackage{mdframed}   

\usepackage{tikz}
\usepackage{tabularx}
\usepackage{ragged2e}

\begin{document}
\title{\hspace{30mm}“Can you speak my dialect?”:\newline A Framework for Server Authentication using Communication Protocol Dialects}
\author{
    \IEEEauthorblockN{Kailash Gogineni\IEEEauthorrefmark{1}, Yongsheng Mei\IEEEauthorrefmark{1}, Guru Venkataramani\IEEEauthorrefmark{1}, Tian Lan\IEEEauthorrefmark{1}}
    \IEEEauthorblockA{\IEEEauthorrefmark{1}The George Washington University
    \\\{kailashg26,ysmei,guruv,tlan\}@gwu.edu}
}


\maketitle

\begin{abstract}
 In today’s world, computer networks have become vulnerable to numerous attacks. In both wireless and wired networks, one of the most common attacks is man-in-the-middle attacks, within which session hijacking, context confusion attacks have been the most attempted. Thus, a potential attacker may have enough time to launch an attack targeting these vulnerabilities (such as rerouting the target request to a malicious server or hijacking the traffic). A viable strategy to solve this problem is, by dynamically changing the system properties, configurations and create unique fingerprints to identify the source (the message was sent from). However, the existing work of fingerprinting mainly focuses on lower-level properties (e.g., IP address, TCP handshake), and only these types of properties are restricted for mutation.\\ 
\indent In this paper, we develop a novel system, called \textcolor{red}{Verify-Pro}, to provide server authentication using communication protocol dialects – that uses a client-server architecture based on network protocols for customizing the communication transactions. For each session, a particular sequence of handshakes will be used as dialects. So, given the context, with the establishment of a one-time username and password, we use the dialects as an authentication mechanism for each request (e.g., \textit{get filename} in FTP) throughout the session, which enforces \textcolor{red}{continuous authentication}. Specifically, we leverage a machine learning approach (pre-trained neural network model) on both client and server machines to trigger a specific dialect that dynamically changes for each request.\\ 
\indent We implement a prototype of Verify-Pro and evaluate its practicality on standard communication protocols FTP (File transfer protocol), HTTP (Hypertext transfer protocol) \& internet of things protocol MQTT (Message queuing telemetry transport). Our experimental results show that by sending misleading information through message packets from an attacker at the application layer, the recipient can identify if the sender is genuine or a spoofed one, with a negligible overhead of 0.536\%.
\end{abstract}

\begin{IEEEkeywords}
Program customization, Protocol dialects, Deep Neural Network, Network security, Authentication.
\end{IEEEkeywords}

\section{Introduction}
\maketitle
\pagestyle{plain}
Communication protocols form the backbone of distributed computing infrastructure, where applications rely on data transfers to execute their tasks. It is, therefore, critical to preserve their security to avoid adversaries from exploiting any loopholes, bugs, and misconfigurations inherently embedded in the relevant software services. Numerous attacks in this threat space have been widely studied in the past- examples include obscuring network sources~\cite{acar2020peek,rupprecht2020imp4gt}, impersonating genuine sites~\cite{delignat2015network,roberts2019you}, Man in the Middle (MITM) attacks through hijacking the request packets~\cite{10.1145/3372297.3417252}, use of proxies~\cite{frolov2020httpt, delignat2015network} and firewalls and replay attacks, where the attackers can easily launch them remotely without establishing a physical connection to their victims. The recent trend in reported attacks
shows a significant increase in threats posed by vulnerabilities in communication protocols~\cite{hashedout,hashedout1}. Apart from that, Barracuda researchers in January 2020 reported that conversation hijacking had increased 400\% in 4 months~\cite{hashedout2}.

In many communication protocols (including the implementation of most popularly used protocols such as FTP~\cite{postel1985file}, HTTP~\cite{dizdarevic2019survey,fielding1999hypertext} \& MQTT~\cite{dizdarevic2019survey,calabretta2018mqtt}), authentication typically occurs prior to the start of the session. This leaves them vulnerable to the above-mentioned attacks, such as MITM-session hijacking, context confusion attacks, where attackers can reroute the request. To counter them, we seek techniques that would ensure \textbf{continuous authentication} for every request in a session through cleverly leveraging \textbf {application layer features}. 

In this work, we present \textbf {Verify-Pro}, a framework that improves server authentication through dynamically and randomly changing the properties of a target (program binary) while maintaining its core functionality of the underlying communication protocol. Furthermore, Verify-Pro dynamically generates customized server replies using protocol dialects for each request, making it difficult for the attacker to launch an effective attack on a constantly self-adapting communication protocol.
In this paper, we define \textit{\textbf{protocol dialect} as variations of protocol implementation at the binary level to incorporate additional security measures. Variations can be in the form of mutating message packets, generating different request-response transactions based on a few environmental conditions, and so on}. 
We provide a proof-of-concept implementation of our proposed tool Verify-Pro using FTP, HTTP \& MQTT protocols. 
\textbf{Our Contribution.} Verify-Pro consists of three major modules: (1) Protocol dialects (PDs), (2) Dialect Decision Mechanism (DDM), and (3) Server Response Verification (SRV). 
The PDs module comprises several customized transactions used for communication between the client and server. 
When a command (e.g., get file.txt in FTP Protocol) is triggered by the client to retrieve a file from the server storage, the DDM module in the client is activated, and the request is fed as input to its neural network and a response dialect `$n$' is determined for future verification. 
We note that the dialect selection must be unpredictable to eavesdroppers in order to prevent the MITM- session hijacking and replay attacks. To this end, we deploy a pre-trained neural network model on both client-server systems equipped with a customized design (by altering the prediction of dialects in a custom-built mechanism) for predicting a dialect number to start the communication. The strategies applied to the neural network are $i)$ \textbf{Uniform distribution of dialects}- offer an advantage in making it hard for the attacker to reverse engineer the neural network (guess the dialect number) as all the dialects are evenly distributed across the sample requests. On the other hand, $ii)$ \textbf{Dialect selection based on cost} property offers a flexible neural network model to trigger the dialect with less cost and make the system more efficient (least cost for a dialect results in that dialect number $D_i$ predicted more frequently across the sample requests), and $iii)$ \textbf{Consolidated loss}- includes a trade-off factor $'a'$ which decides the sensitivity to the above-described properties. To our knowledge, this paper is the first to use a neural network as a decision mechanism (DDM) in triggering the dialects that dynamically changes the transactions for each request. The general and straightforward approach for the decision mechanism at hand is directly using a "real" pre-shared key that could be used as an index in selecting the dialect from the dialect library. In contrast to this approach, verify-pro, $1)$ proposes a decision mechanism with custom-built properties to design a customized neural network model for the prediction of dialect, $2)$ Less deployment cost \& minimal overhead, $3)$ the use pre-shared key (secret key) is vulnerable to brute-force attacks~\cite{psk}, whereas reverse engineering a neural network is comparatively harder to inverse and these attacks can also be mitigated~\cite{liu2019mitigating, wang2019npufort}.   
Since the client needs to verify the server's dialect in which the response was dispatched, the SRV module on the client side verifies if the server responds to the request using the `correct' dialect `$n$'.\\ 
\textbf{Contributions}. To sum up, we make the following contributions:
\begin{itemize}
\item We propose Verify-Pro, an automated framework for applying communication protocol dialects as fingerprints to authenticate servers. We harness different protocol dialect implementations and leverage them to create unique responses that help authenticate servers during communication and improve security.
\item  Verify-Pro uses a neural network model to select a unique dialect for response to be used for each request. The motive behind the Neural network model is to deploy a customized mechanism for the selection of dialect and avoid reverse engineering attacks by adversaries. 
\item  We design and implement Verify-Pro prototype on  FTP, HTTP \& MQTT, and evaluate its effectiveness using dialects as fingerprints on a real-world setup. Our evaluation results show that Verify-Pro can successfully defeat MITM-Session hijacking, replay \& context confusion attacks.
    
\end{itemize}
\textbf{Motivation.} Communication protocols face an increasing number of attacks such as proxies and session hijacking attacks-MITM, which become the main threat to communication security. In this paper, we create, analyze and evaluate the strategies that are diverse and that repeatedly shift and change over time to increase the complexity, time, and cost for the attackers, thereby increasing the system resiliency. Existing fingerprinting methods are limited to increasing complexity against potential attacks because existed network system properties (e.g., IP address, TCP three-way handshakes, port numbers) are confined for mutation. Lack of additional authentication to validate the identity in the session benefits the MITM attackers to reroute the target requests to malicious entities-resulting in malformed messages. Forestalling this vulnerability can happen by enforcing \textbf{continuous authentication} for each request.\\
To this end, we design a resource-constrained (e.g., low cost) tool, Verify-Pro, which creates a number of variations by using protocol dialects as fingerprints that leverage more \textbf{application layer features}. Protocol dialects can increase the variations of system properties to improve the security of communication. We design a Dialect decision module (DDM) to ensure the transmutation of dialects used for every transaction. Verify-Pro allows the client and server, who are aware of the same protocol dialect varying patterns, to communicate with each other accurately. In addition, using a data-driven decision tree model (SRV) prevents the overlapping of responses sent from different dialects or packets sent by an attacker.\\ 
Our experimental results demonstrate that the techniques we used successfully identify the (and also rejecting– close the connection of the entire session) attackers tending to send unwanted or malformed messages. Moreover, we show that after we successfully identify the message is from an attacker, we close the connection of the entire session, which helps the recipient from receiving subsequent message packets (responses) from the attackers. We believe that extracting the DDM module and adapting to dialect evolution comes at a high cost and makes the attacker invest more time.\\
\textbf{Organization.} The rest of this paper is organized as follows: Section I provides Introduction \& Motivation. Section II discusses the Background of communication protocols. Section III discusses the Threat model \& Assumptions. Section IV discusses the System Overview. Section V discusses the Implementation. Section VI discusses security analysis. Section VII discusses our experiment evaluation and usability. Section VIII discusses the related work. The last section presents the future scope with a conclusion on the experiment performed.

\section{Background}
The File Transfer Protocol (FTP), as defined in RFC 959 \cite{postel1985file}, is a text-based protocol that is widely used for transferring files between computers over a TCP/IP network, such as the internet. FTP promotes the exchange of data between the people within their offices and across the internet. FTP is built on a client-server architecture using separate control and data connections between the client and the server. The FTP has a default authentication mechanism such as the username and password before initiating the session. Messages sent by the sender are called requests (e.g., get file.txt), and they instruct the receiver to transfer the data based on the request. The FTP RFC defines ten commands. Each command has unique usage and different syntax to be sent as a request. When the sender types a request (get hello.txt), the control connection is used to transfer these commands to the server, and once the receiver reads the request, files are transferred only via the data connection if the file is present in the server storage. The data is typically transmitted as a stream of bytes over the data channel to the client. In this paper, we solely focus on the GET command, where a client’s request format is:
\begin{center} \[
\small {Request = \{ Command\quad Argument \in get\quad filename \}}
\]
\end{center}
In this FTP implementation, a command and an argument are separated by a “space”. The first word that occurs before space is treated as the command, and the subsequent one is the filename of the file to be retrieved. We refer to various implementations on how the FTP client-server systems communicate. As we will discuss these later, we leverage these deviations to determine the FTP dialect spoken in specific FTP conversations. The background details of HTTP \& MQTT protocols are described in \S ~\textcolor{red}{\ref{eval}}.\\
\vspace{-\baselineskip}
\section{Threat model}
From an offensive perspective, the attacker's objective is to gather relevant information used to send an unwanted or malformed response to a target machine. 
Verify-Pro is a server authentication mechanism designed for ease of use while providing continuous authentication for each request (e.g., \textit{get filename} in FTP) in the session. In addition, it allows the client to verify the authenticity of the server's responses (in each transaction) over untrusted networks. Verify-Pro is intended to protect against the following types of adversaries:
\begin{enumerate}
    \item \textbf{Active adversary.} We allow the attacker to eavesdrop on any network traffic passively. An active adversary can actively divert the requests and responses exchanged between the client and server machines. For example, the active adversary can replay, use proxies, intercept, fabricate new messages and stop messages from reaching their destination (by sending the request to a malicious server)-request hijacking.
    \item\textbf{Context Confusion attacks.} We assume the attacker locates in the same LAN as the victims, being able to reroute the encrypted traffic (request), where the MITM attacks rely on shared TLS certificates~\cite{alashwali2018s}. Using the shared TLS certificates~\cite{10.1145/3372297.3417252} adversaries can cause origin confusion issues~\cite{10.1145/2736277.2741089}. Intrinsically, we assume the typical adversaries sniffer in local Wi-Fi and locate in the open Wi-Fi network without strong security protection. Also, the attacks can be launched by gateways or proxies.
\end{enumerate}
Mentioned threat models support a proof-of-concept for Verify-Pro as a promising new direction to use dialects as fingerprints for providing server authentication between the client-server systems in communication security. 
We make the following assumptions to support the Verify-Pro system:\\ 
    $\bullet$ We provide one-way authentication, where two parties communicate; only one party will authenticate to the other entity. In our case, as we assume the client is ideally genuine, the server needs to authenticate itself to the client. The responses sent from server to client can be malformed or replayed, and the request sent by the genuine client can be hijacked to a flawed server.\\
    $\bullet$ We further assume that the attacker has no means to access the dialect decision model (a deep neural network that decides a dialect), directly compromise the software, storage, and data structures executing on the client and server.\\
While this threat model under-estimates the scope and severity of some active and future day attacks against communication protocol, it supports the proof-of-concept for protocol dialects as a promising direction to add continuous authentication for each request in the session along with default username and password mechanisms. Hence, this work functions as an initial step directing a comprehensive solution based on using communication protocol dialects as fingerprints.  
\begin{figure}[t]
\centering
\includegraphics[width=8.5cm]{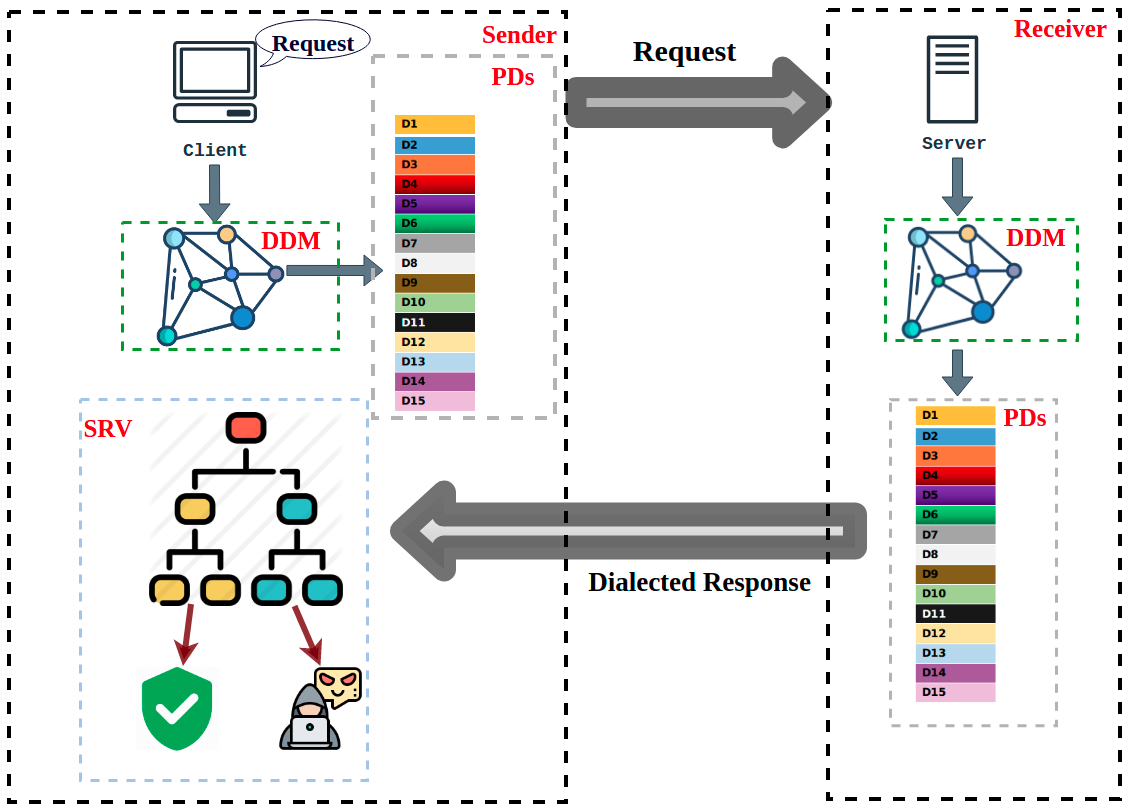}
\caption{Verify-Pro System Diagram.}
\label{Figure1}
\end{figure}

\section{System Overview}
Verify-Pro consists of three major modules: (1) Protocol dialects (PDs), (2) Dialect Decision Mechanism (DDM), and (3) Server Response Verification (SRV). In Figure~\textcolor{red}{\ref{Figure1}}, we provide an illustration of the Verify-Pro system diagram. \\
The attacker can actively try to intercept the networks by performing session hijacking and context confusion attacks by placing the realistic MITM attackers to reroute the target message on a wireless network. To counter these attacks, the main aspect of our proposal is to utilize the protocols dialects (PDs) (distributed to client-server systems) as an effective server authentication system. These dialects are embedded so that the client has request packets, and the server includes the response packets. In contrast to a cryptographic real shared key, we use a custom-designed DDM module (with three properties to build a custom neural network model which gives a result of divergence in the prediction of dialects) to select the dialect for initiating the communication. During the actual communication, various dialects will be selected dynamically by the DDM module, and this, in turn, gives a better edge over the default implementations of communication protocols as the request-responses keep dynamically changing for every transaction and confuses the middle attackers, as a result of enforcing our continuous authentication. When a request packet (e.g., get file.txt in FTP) is triggered by the client to retrieve a file from the server storage, the DDM module in the client is activated, and the request is fed as input to its neural network to obtain a specific dialect number in which it can send its request to the server and a response dialect ‘$n$’ is determined for future verification. We note that the command request (e.g., get file.txt in FTP) in all dialects is in the same format, and also, our protocol dialect-ing actually starts from the server’s response. 
Note that the neural network model used on client-server systems is the same, so given the context, for the request “$R_{i}$”, both the client and server should predict the same dialect number “$D_{i}$”. Followed by the prediction of dialect number by the server, a dialect-ed response “$Resp$” is dispatched to the client. The SRV module on the client side verifies if the server responds to the request using the “$correct$” dialect “$n$”. Verify-Pro has the protocols dialects deployed in a decentralized manner that utilizes the DDM module to trigger the dialect to validate the server's response. To pass the validation, the client-server systems must share the same neural network model \& protocol dialects in an automated manner. The server responds to the request sent by the client in a unique way for each dialect. To launch a successful attack, the attacker has to spoof the responses sent to the client or forward the requests to a flawed server. This is an impossible task for the MITM attackers as we are enforcing continuous authentication for each request in the session. The core idea of Verify-pro is that, even if the attacker passively observes the traffic, controls the wireless networks, it cannot predict the dialect number to reply or start the communication as we import dialect library \& DDM module on client-server machines with custom-built dialects and prediction schemes.

\subsection{Protocol Dialects}
\label{section3.2}
This section describes the process of creating a dialect for a protocol and integrating it into an existing binary of the communication protocol. The PDs module summarizes the resources, rules utilized to ensure that the protocol dialects will meet the requirements of communication protocols for imposing continuous authentication. The RFCs that define communication protocols delineate the protocol that a server has to speak to properly communicate with the client. However, different client-server systems might implement the communication protocol in slightly different ways, for three main reasons:\\
1. The RFCs do not always provide a single possible format when specifying the request-response packets. For example, in FTP Protocol command identifiers are case insensitive and also normal binary source code variations, which means the $GET$ or $get$ are both valid commands. Since the client has $get$ request format - "$get\ \ filename$", we could even modify the program binary to change the placement of command and filename arguments such as "$filename\ \ get$". \\  
2. We identify numerous Communication protocol implementations and cross-graft the variations into standard protocol binary. At the core of creating customized protocol dialects, the critical problem is to understand the variations in the handshakes. Therefore, we refer to various implementations of the communication protocols which helps us to generate customized transactions which can be used as dialects.\\ 
3. Communication Protocols like FTP have no headers and fields at the application layer. Therefore, any client-server system which performs the basic core functionality of effectively sharing the file is considered as the core functionality of FTP. We take the advantage of this scenario by manually creating response packets with fields separated by a comma -“,”. Apart from that, protocols like MQTT, HTTP have headers but still we show that our Verify-Pro is efficiently compatible with those protocols in $\S$~\textcolor{red}{\ref{eval}} \& Appendix \textcolor{red}{\ref{appendix:c}}.\\
In this paper, we call different implementations of the network protocols, protocol dialects. 
Given the context of a standard communication protocol, \textit{a dialect $D$ is defined as the variation created by the cross-grafting various implementations of handshake rules and mutating the message packets while keeping the core functionality of the protocol intact. More precisely, we assume {$D_{1}$, $D_{2}$, $D_{3}$,….$D_{n}$} are the dialects created by using different communication rules and mutating the request-response packets. If a user initiates a request “$R_{i}$", a dialect “$D_{i}$” will be used for communication. Thus, each protocol dialect “$D_{i}$” is uniquely defined by a request “$R_{i}$” by using a DDM module.}
\subsubsection{Message templates}
\begin{figure}[t]
\centering
\includegraphics[width=9cm]{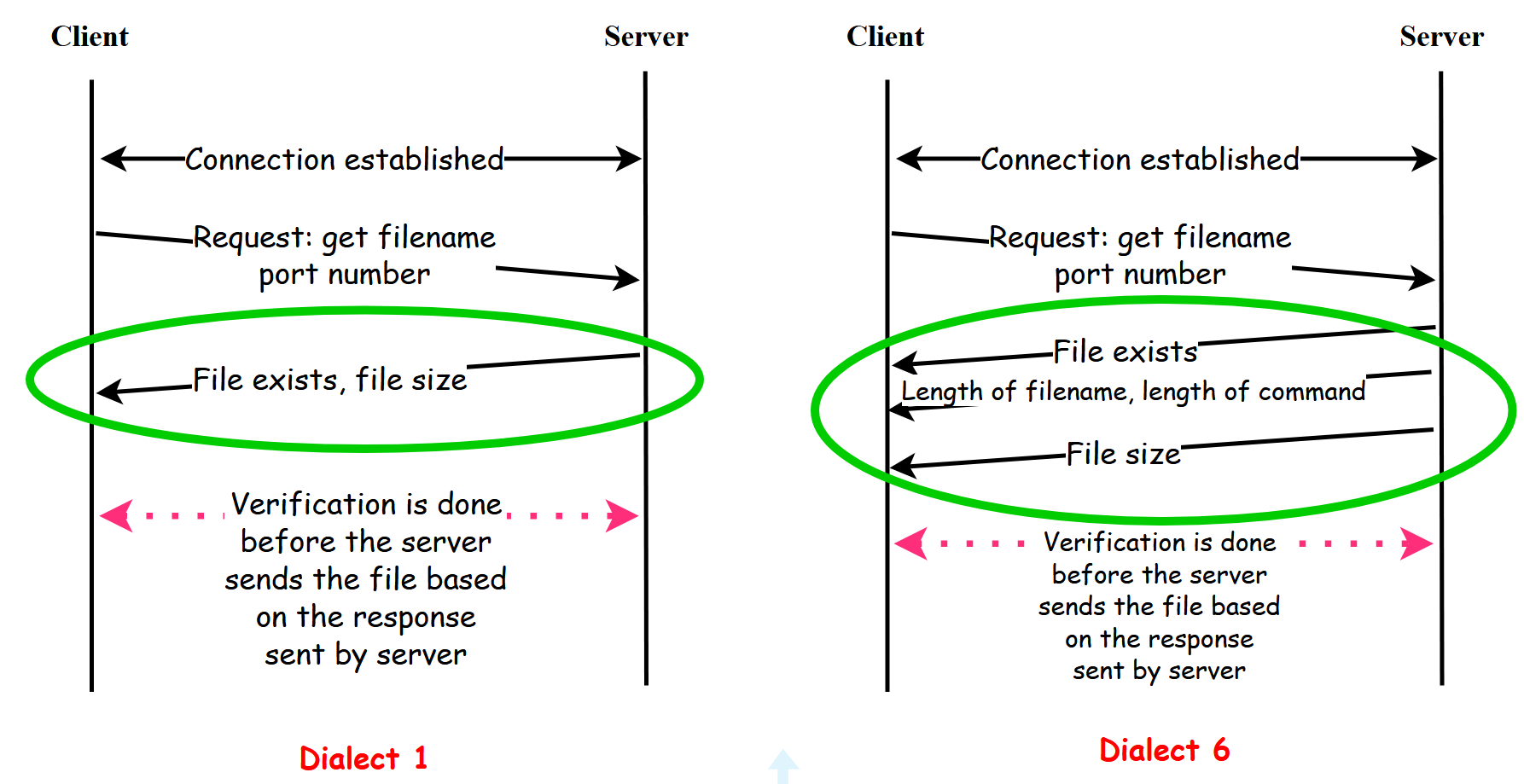}
\caption{Request-Response of Dialect 1 \& Dialect 6 in FTP.}
\label{Figure2}
\end{figure}

As discussed in the previous section, we only provide one-way authentication, meaning the server’s response should prove its identity to a genuine client. We create the message templates for customizing the server’s response. As the communication protocols do not have any format for messages, we define fields as the delimiters, and they are separated by a comma. We note that the information of (command, length of command, length of the filename, filename, file size, etc.) (For space reasons, \textit{FTP protocol dialects are listed in Appendix \textcolor{red}{\ref{appendix:A}})} are used for defining a specific dialect so that the client knows the response (dialect pattern) it will receive beforehand if both of them communicate in identical dialect. As the communication protocols comprise request-response architecture, we also tweak the communication by including more number of request-responses in a single dialect creation. We show the difference between the default transaction and customized transaction for FTP protocol in Appendix \textcolor{red}{\ref{appendix:b}} and the handshake mechanism of model dialects for FTP protocol in Figure \textcolor{red}{\ref{Figure2}}. 
The server's response to different dialects is highlighted in green. Each has a unique message structure that helps the client identify the dialect number which the server used to send its response. 
\subsubsection{Message format variations}
These variations represent transmutations in the format of replies that the server sends back to the client. For instance, the FTP server replies to a client’s command have an unconventional structure, i.e., each dialect has its server’s response structure uniquely, and the response is provided as human-readable information to the user. Following protocol specifications, a client should read the data from the server until it receives a line terminator, check the message, and forward the text of the reply to the user if required. Considering the protocol specifications, we craft the reply variants in two different ways to systematically explore how a client identifies them:\\
\textbf{Non-compliant replies:} These reply variations do not comply with the protocol specification, but are frequently observed in the conversations. For instance, this technique might vary the messages sent to the client, vary the capitalization of the reply (upper/lower/mixed-cases) including unconventional messages such as - "$Yes, file\ \ exists$". Techniques such as unconventional handshakes, packet mutation, field shifting (field altering for command name, command length, length of filename, etc.) are integrated to increase the software diversity in the program  binaries of network protocols.\\ 
\textbf{Replies with fields:} This technique might vary the number of messages sent in a particular packet with fields controlling them. In this scenario, the server sends the information separated by a comma in the same packet. Of course, it is not guaranteed that the client requires information such as command name, length of the filename, file size, etc. We use this information to make the genuine server send a response in such a way that the genuine client can understand its response structure. We do not use an automation technique to deploy the dialects because the dialects are designed in an unconventional structure (such as using message fields, packet \& handshake sequence mutations- traced from various protocol implementations). The protocol dialects are spawned as different versions of a protocol deployed into the single communication protocol binary. Furthermore, deploying dialects as threads gives us an added advantage of minimal overhead and less cost, as the triggering of each dialect happens in milliseconds. Although for future work, we aim to add more dialects to increase the complexity and make the system robust with less cost and less execution overhead.
\subsection{Dialect Decision Mechanism}
Dialect selection by the client and server must be unpredictable to eavesdroppers, and yet ensure that both client and server systems predict identical dialect $'D_i'$ for a given request $'R_i'$. To this end, we deploy a pre-trained neural network model which confers a flexible \& customized neural network with certain properties. The details of DDM module dataset are described in Appendix \textcolor{red}{\ref{appendix:a.1}}.\\ 
\textbf{Training process.} Our model is a simple DNN, which is able to map the input feature vectors $x$ = ${x_1, . . . x_n}$ \textit{(converting the “request” of the protocol into vectors)} consisting of $n$ samples to an output $y_i$ \textit{(which is the dialect number for a given request)}. The input of the neural network has a size of $n=100$ \textit{(vector for each request)}, fed as a high dimensional feature vector. The model has two hidden layers with 128 neurons each and “relu” activation function in each layer but the last layer has 15 neurons \textit{(this represents we have n = 15 as the total number of dialects for FTP protocol)} with “softmax” activation function. An increase in the number of layers proved to produce good results and make the trained model complex. The ADAM optimizer was used for the training process. The models were implemented by using Python and Keras with Tensorflow backend. 
The entire layer-wise computation procedure from input data samples to output prediction is defined as inference. When ground truth data is not available, we consider the loss function $L$ to be imposed as a criterion that can be useful in the generation of labels (dialect numbers). The loss function of the model consists of three variants: \\
\textbf{Uniform distribution of dialects.} This property implies that if we have a set of input requests in the testing dataset, the predicted dialect numbers should be uniformly distributed.  The testing dataset has $K$ inputs, then expected dialects should be $K/N$, where is $N$ is the number of dialects. To achieve this property, we used entropy maximization in the loss function for training. Here, the $P(y_i)$ represents the occurrence of particular dialect number of request $y_i$. (computed as the number of occurrences of requests with a particular probability $i$ divided by the number of all requests of that particular family), log2 is a logarithm with base 2, and $M$ is the total number of dialects (classes). This property makes the neural network model resilient as the attacker will have to invest time and effort to predict or inverse the model as all the dialects have an equal probability of occurring.
\begin{equation} \label{eq1}
\small    Uniformity\ \ loss(l1): \ min \sum_{i=1}^{M}(P(y_i)log_2(P(y_i))) \   
\end{equation}
\textbf{Cost as a criteria for dialect selection.}
We define the cost of dialect as the number of request-responses (the number (cost) is multiplied by 1000 to achieve significant variations) shared between the client and server machines of a particular dialect. We assume cost $C_x$ for each dialect and $P(y_i)$ is the probability distribution of choosing a dialect, then we aim to minimize the sum of $P(y_i)\times C_x$ which is the expected cost. $M$ is defined as the total number of dialects (classes). This property offers the flexibility to make custom predictions based on the cost individually assigned to each dialect. For example, we intended that dialect 4 needs to have a high chance of prediction, whereas dialect 8 should have the least chance. In this case, we assign a higher value as the cost to dialect 4, whereas the least number (cost) for dialect 8. After training, the model would predict dialect 4 with a high frequency. Customization with cost makes the system more flexible to revise the prediction frequently and confuse the MITM attackers. 
\begin{equation} \label{eq1}
\small   Cost\ \ based\ \ dialect\ \ loss(l2): \ min \sum_{i=1}^{M}(P(y_i)(C_x)) \   
\end{equation}
\textbf{Consolidated loss.} We use the formula from equation \textcolor{red}{\ref{eq3}} to calculate the consolidated loss using a trade-off factor ‘$a$’, in range [0,1]. The combination of these losses l1 \& l2 proved to be effective such as, by varying the ‘$a$’ value, the prediction of dialects will gradually change from ‘finding the dialect with low cost’ to ‘evenly distributing the dialects’ across the sample requests and allows a customized design for prediction of dialects.
\begin{equation} \label{eq3}
\small   Consolidated\ \ loss(l3):  (a \times l2) + ((1-a) \times l1)  \
\end{equation}
The algorithm of Dialect Decision mechanism is shown in Algorithm \textcolor{red}{\ref{alg:DDM}}. It starts with the user inputting a request (e.g., get file.txt) $'r'$. Then, request $'r'$ is fed as input to the DDM module on the client system, and a response dialect $'n'$ is predicted. Now, the undialect-ed request from that particular dialect instance is sent to the server. The convergence graphs for the properties are shown in Figure~\textcolor{red}{\ref{trade-off1}}. Finally, the client also computes the dialect number to utilize it for future verification by the SRV module in Algorithm \textcolor{red}{\ref{alg:SRV2}}. 

\begin{algorithm}
\caption{Dialect Decision Module algorithm - DDM}
\label{alg:DDM}
\textbf{Input}: R: A request R (e.g., $'get\ \ filename'\gets R$ to retrieve a file)\\
\textbf{Output}: D: A dialect number from a dialect pool of range 1 to 15.
\begin{algorithmic}[1]
\Procedure{DDM}{$R$}\Comment{Prediction of Dialect D from a request R}
\State $R\gets get\ \ filename$ \Comment{Request is triggered, get filename}
\State $D\gets Neural\ \ network\ \ model(R)$ \Comment{Request is sent as input to NN}
\State \textbf{return} $D$\Comment{The Dialect number is D}
\State $Class\_\ D \gets D$ \Comment{Code block of a dialect}
\State $R \gets Class\_\ D$ \Comment{R is undialected request}
\State R is sent to Server machine listening on port 21
\EndProcedure
\end{algorithmic}
\end{algorithm}
\vspace{-\baselineskip}
\subsection{Server Response Verification}
After receiving the server's response, the client sends the response as the input to the decision tree, which verifies the response structure sent by the server.\\
The decision tree is one of the most commonly used Classification algorithms in machine learning. Decision trees have good explainability and are widely used, interpretable models. Hence, they are deployed in applications such as product recommendations, fraud detection, and Intrusion Detection. Moreover, a decision tree provides unique perceptiveness in identifying malicious activity and can assist in systematic analysis. Decision trees are used to identify \& organize conditions with rules in a structured process.  \\
\textbf{Feature selection.} In a typical supervised learning scheme, we are given a learning algorithm $\delta$ and training set $T \subset \alpha \times \beta$ comprised of elements of some set $\alpha$, with its classification label from a finite set of classes $\beta$. Applying $\delta$ to $T$ results in a classifier model: $\eta: \alpha \times \beta$. Pertaining with our mechanism, we define $\alpha$ as response structure (each dialect has a different response structure, for instance, server's responses in FTP protocol for numerous dialects are shown in Appendix \textcolor{red}{\ref{appendix:A}} (highlighted in red color)), $\alpha_{i}$ are the features, and $\beta$ or $class$ is the dialect number. We assume that Decision tree: $\tau$ is equipped with a finite set of features with which it can construct a decision tree classifier. In the response structure, we consider each packet has utmost 4 fields and there are no more than 6 packets in the response structure. Features like number of packets, number of fields in each packet, data type of each field are taken into consideration for the feature set. Ideally, the learning algorithm $\delta$ would consider every possible sequence of features to partition a dataset to arrive at an optimal classifier. Dataset details are described in appx.~\textcolor{red}{\ref{appendix:a.1}}.\\
In a decision tree $\Delta$, each intermediate node includes an attribute, each branch corresponds to a decision and each leaf node indicates an outcome. More precisely, each internal node $n$ has a corresponding attribute index $n.attribute$ from the feature set of $d$ attributes, and a threshold $n.threshold$ and two children $n.left$ and $n.right$. Each leaf node $y$ comprises of the classification result $y.class$, which is the label. Each request is depicted as a sized vector $R$ relating to each attribute. The algorithm of decision tree prediction is shown in Algorithm \textcolor{red}{\ref{alg:SRV1}}. The mechanism initiates from the root node of $\Delta$. For each node of $n$ in $\Delta$, it compares $R[n.attribute]$ with $n.threshold$, and decides to go to $n.left$ if $R[n.attribute]$ $<$ $n.threshold$, and $n.right$ in different circumstances. In the end, the algorithm reaches a leaf node $y$ and the outcome of the prediction is $y.class$ (dialect).\\
We use the sklearn package in Python, which assists us in designing a decision tree without manually specifying the rules for classification. Sklearn uses CART~\cite{breiman1984classification} (Classification and Regression Trees), which constructs binary trees; specifically, each internal node has two outgoing edges. To train a CART decision tree classifier, given a training dataset, the decision tree is obtained by splitting the set into subsets from the root node to the children node. The splitting is based on the rules derived by the Gini index (equation \textcolor{red}{\ref{eq1}}), and the splitting process is repeatedly performed on each derived subset in a recursive manner, known as recursive partitioning. In our scheme, we only consider the pre-trained decision tree model on the client-side, confirming that the sender's response is from the ‘correct’ dialect $n$, that the client was expecting.
\begin{algorithm}
\caption{Decision tree prediction for sample request $'R'$}
\label{alg:SRV1}
\begin{flushleft}
\textbf{Input}: Decision tree $\Delta$, sample request \textbf{R}\\
\textbf{Output}: Classification result: Dialect number $D$ for a request $R$ 
\end{flushleft}
\begin{algorithmic}[1]
\State $n$ = $\Delta$.root
\While{$n$ is not a leaf node}
\State \textbf{if \small R}[$n.attribute$] $<$ $n.threshold$ \textbf{then}
\State \      \ $n$ = $n.left$
\State \textbf{else}
\State \      \ $n$ = $n.right$
\EndWhile
\State return $n.dialect\_number$ \Comment{dialect\_number is the class}
\end{algorithmic}
\end{algorithm}
\begin{equation} \label{eq1}
\textbf{Gini index:}
    \ I_G = 1 - \sum_{i=1}^{c}p^2_i \   
\end{equation}
where $P_i$ is the proportion of samples that belongs to a class (dialect) $c$ for a particular node. In algorithm \textcolor{red}{\ref{alg:SRV1}}, the threshold is based on impurity metric, Gini index, which is chosen by the sklearn package itself.
\begin{algorithm}
\caption{Server Response Validation algorithm - SRV}
\label{alg:SRV2}
\begin{flushleft}
\textbf{Input}: R: A request R (e.g., $'get\ \ filename'\gets R$) is received from the client\\
\textbf{Output}: Response Validation by the client.
\end{flushleft}
\begin{algorithmic}[1]
\Procedure{Server side: DDM}{$R$} \Comment{Prediction of Dialect D from a request R}
\State $R\gets get\ \ filename$ \Comment{Request is triggered, get filename}
\State $D\gets Neural\ \ network\ \ model(R)$ \Comment{Request is sent as input to NN}
\State \textbf{return} $D$\Comment{The Dialect number is D}
\State $Class\_\ D \gets D$ \Comment{Code block of a dialect}
\State $Resp \gets Class\_\ D$ \Comment{Resp is dialected response}
\EndProcedure
\State Resp is sent to Client machine.
\Procedure{Client side: SRV}{$Resp$}
\State $Client \gets Resp$ \Comment{The response is sent to client}
\State Client: $DS \_\ tree(Resp)\ \  $ \Comment{Client verifies the server's response by matching response structure using a decision tree and also verifying the contents} 
\State \textbf{If} Resp $\in $ Client's expected dialect's response \textbf{then}
\State $\bar{Resp}$ from the client is sent to server
\State \textbf{Else} Connection is terminated.

\EndProcedure
\end{algorithmic}
\end{algorithm}\\
\textbf{Decision tree based response verification.} In our decision tree model, an input $x$ is traversed in the tree learned on $\delta$. We deploy a pre-trained decision tree model on the client-side to check the overlapping (pattern) of the response sent by the server. For example in FTP protocol, we show our server response - the input is:
\begin{center}
    Input: “\textit{P1: command/ P2: filename, length of filename}"
\end{center}
For the learning purpose, we convert the input into sized vectors, and we consider the data types with fields separated by “,” and the packets separated by “/". Packet 1(P1) has a string as the first field for the above input, and packet 2(P2) has a string as the first field and an integer as the second field. Since we verify the data type of each field in the packet and the structure of the packet, this compels us to create a data set of 150K samples with random strings and integers according to each dialect response pattern. Following the vector conversion, the input is traversed through the pre-trained model, and a class (dialect number) is predicted with which that response structure dwells. Likewise, the decision tree performs extremely well in verifying the response structure as all the dialects, and their responses are unique. The algorithm of Server Response Validation is shown in Algorithm \textcolor{red}{\ref{alg:SRV2}}. It starts with the server receiving the request sent by the client; again, the request (e.g., get file.txt) is fed as input to the DDM module on the server-side, and a response dialect $'n'$ is determined. Based on the dialect number, the dialect-ed response $'resp'$ is sent to the client. Eventually, when the client receives a dialect-ed response $'resp'$, the SRV module gets activated, and the $'resp'$ is fed as the input to the SRV (Decision tree) module, which assists the client in verifying that if the response was actually from the dialect $'n'$, the client was expecting/is there any overlap of dialects/ variations in packet structures (occur because of the middle attacker or flawed server's). After verifying the message pattern received and confirming the dialect, we also verify the values of fields, as the client already knows some information about the message it will receive (such as command, filenames, length of command \& filenames).
\vspace{-\baselineskip}
\section{Implementation}
\label{Implementation}
We implement a prototype of Verify-Pro on communication protocols in three phases, which include the following main modules.\\
\textbf{Phase 1: Protocol dialects (PDs)}: We implement our customized transaction functions (e.g., mutating the message format variations) in the communication protocol. Handshakes and message formats are cross-grafted from different sources of Network Protocol implementations, and some according to environmental conditions. We leverage these handshakes and use them as fingerprints to verify the identity of the response sender. We design 15 dialects (in FTP) as a proof-of-concept, and they are deployed in both client and server machines to communicate effectively in one of the dialects triggered for each request.\\
\textbf{Phase 2: Dialect Decision Mechanism (DDM)}: We implement the DDM module as a deep neural network which has input as the $'request'$ (e.g., \textit{get file.txt} in FTP protocol), label as the dialect number (ranges from 1 to 15). The output of the DDM will be used as the dialect number to start the communication, and the customized handshake is initiated.\\
\textbf{Phase 3: Server Response Verification (SRV)}: We implement our SRV module to verify the structure (format in which the packets are sent) of the sender's response (i.e., server's response) to avoid overlapping with any dialect's response or any malicious response. 
The SRV module will have a decision tree to validate the structure of response packets received by the client. Provided, if the client confirms that the response was from the server’s dialect as - \textbf{correct}, that it was actually expecting from, then the communication will be successful. Any deviation from this process results in the termination of the entire session.

\section{Security analysis}
\label{seecurity}
This section highlights the security risks of the Communication protocols and show how they can be prevented by Verify-Pro. We summarized our findings in Table \textcolor{red}{\ref{table1}}.\\
\textbf{Attack description.} Attack vectors such as MITM-session hijacking- rerouting the target request~\cite{chordiya2018man,antikainen2014spook}, context confusion attacks, use of proxies/firewalls can intentionally pretend to be a legitimate user, which damages the authentication of communication. The rerouting of target request from a genuine client to a flawed server by the middle attackers, cause meddling in the data channel. In this scenario, the adversary is capable of setting up a MITM relay or fake base station~\cite{wong20man} between the communication parties to disrupt the communication, through which the genuine server is renounced to receive the target request. Context confusion attacks~\cite{10.1145/3372297.3417252}, used by standard protocols like HTTPS, FTPS, a standard mechanism such as SSL/TLS can be used to protect the communication channel. Nevertheless, aiming at the vulnerability that existed in SSL/TLS, the middle attackers can launch cipher suite downgrade attacks by abandoning the \textit{Clienthello} packet sent by the genuine client and replacing it with a malformed packet consisting lower version of SSL/TLS~\cite{alashwali2018s,MITM1,lee2020return,sjoholmsierchio2020strengthening}. In the context confusion attack, the attack occurs because of shared TLS certificates; thereby, the adversary reroutes the request is bypassed to a flawed certificate-sharing server. Use of intermediate entities like malicious proxies~\cite{de2016killed,durumeric2017security,chung2016tunneling,soghoian2011certified,kondracki2020meddling,frolov2020httpt} or interception software's~\cite{de2016killed,durumeric2017security}, a form of MITM attacks, cause the origin confusion issues as the attacker can hijack the secure traffic between the client and server. \\ 
\vspace{-\baselineskip}
\subsection{Attack setup in the Real world}
\begin{figure}[]
\centering
\includegraphics[width=9cm]{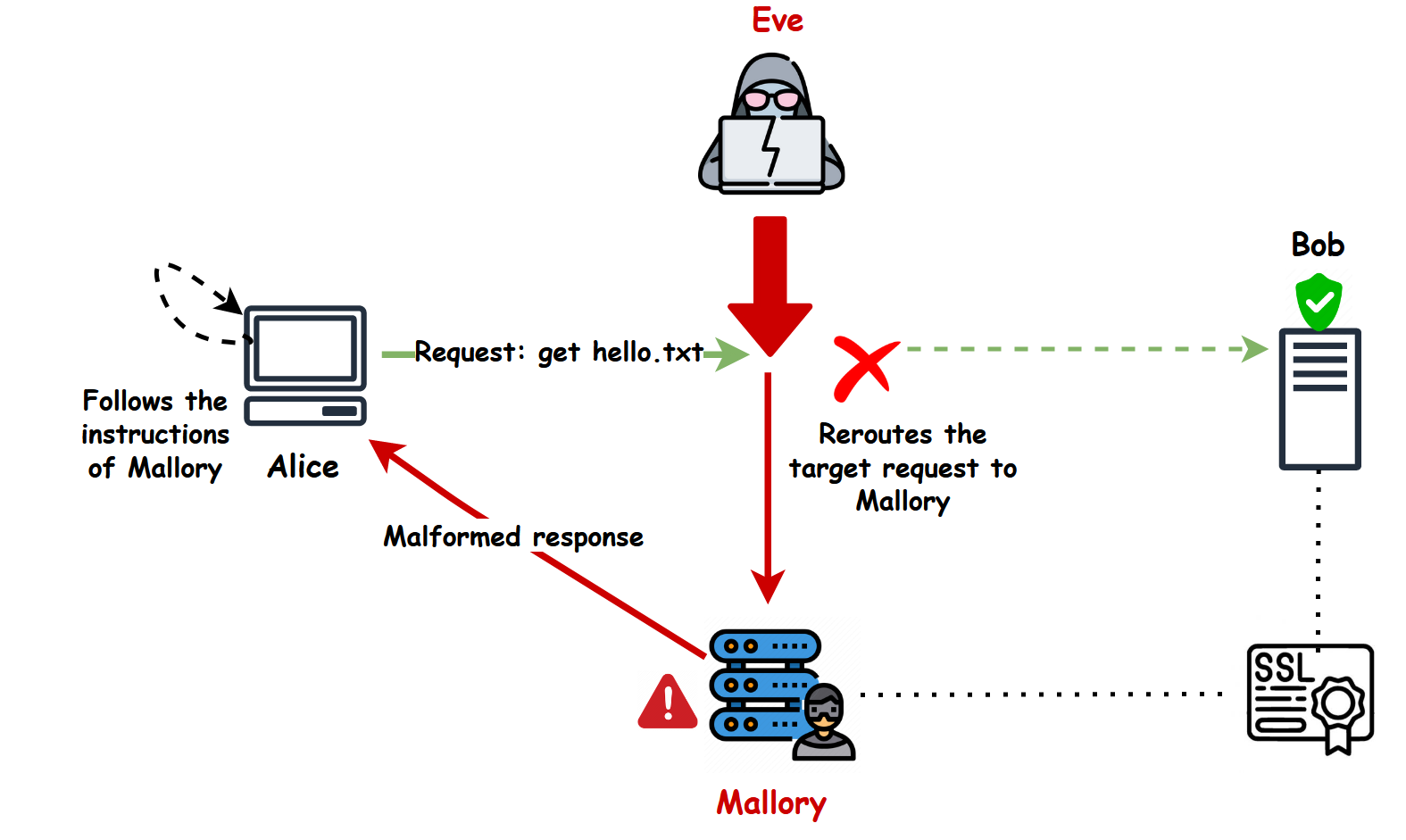}
\caption{Attack setup.}
\label{setup1}
\end{figure}

We introduce the scenario of bypassing the security policies and demonstrate the context confusion, MITM-session hijacking attacks in a complex environment. To understand the real impacts, we conducted a real-world experiment for this type of data forwarding attack (Figure \textcolor{red}{\ref{setup1}}). Verify-Pro requires a genuine client (Alice), genuine server (Bob), MITM attacker (Eve) \& attacker server (Mallory) components. $Eve$ middles himself between the two communicating entities $Alice$ and $Bob$. $Eve$ then captures the target request and forwards the request to $Mallory$, which is possible through the spoofing attacks~\cite{port, 8500144}. $Mallory$ successfully impersonates himself as $Bob$ and communicates with $Alice$ after the target request is received by $Eve$. This situation arises when both the $Bob$ \& $Mallory$ share one valid TLS certificate, and the authentication also can be passed because of the same TLS certificate~\cite{10.1145/3372297.3417252}. Therefore, the data forwarding route remains valid, which does not create any uncertainty for $Alice$. Thus, $Mallory$ can communicate with $Alice$ and stay on the communication link as long as possible by delivering the traffic normally. In the end, $Mallory$ starts the communication in weak security policies (because of TLS downgrade attacks) with $Alice$, resulting in the failure of authentication. As such, $Mallory$ sends the malformed response (without knowing any dialect pattern), resulting in authentication issues that can occur at any target request and may cause potential security threats. \\
A defense that successfully prevents this vulnerability is by enforcing continuous authentication using protocol dialects (by leveraging application layer features) during communication. In making our security claim, we send a request $get\ \ file.txt$ from $Alice$ to the DDM module to predict a dialect, say $D8$. Recall that the $request$ of every dialect is of the same format and the dialect-ing actually starts from the $Bob's$ $response$. From the above threat models, we presume that when the target request is hijacked and diverted by a $Eve$ to $Mallory$. Since $Bob$ \& $Mallory$ share the same TLS certificates, there exists no way to validate the identity of response packets. Under these assumptions, we claim that the $Alice$ expects a response in \textit{single packet containing four fields} from $D8$. To support this, we use the PDs module (which comprises of unique protocol dialect \& unique response structures for each dialect $D_i$) \& DDM module to select a dialect based on the $Alice's$ request $R_i$. In turn, $Alice$ uses the SRV module to verify \textit{if the response has a single packet with four fields} (separated by $','$). To this end, $Alice$ has the features of all the dialects such as the number of packets, fields, and data type of each field. To further make this verification strong, $Alice$ already knows the message pattern \& some information (such as command name, filename, length of command) of the dialect $D8$ in advance. In this way, all the dialects are equipped with unique features which assist the client - $Alice$ in identifying their identity. In this case, the adversary- $Mallory$ should create a shadow model for the DDM module and PDs in convincing $Alice$ to complete the handshake. However, we argue that adapting to these modules incurs time \& cost complexity to the attacker- $Mallory$. For each request $R_i$, the responses $Resp$ change dynamically because of the DDM module. Specifically, if $Mallory$ sends a packet that does not contain the communication rules induced by PDs, $Alice$ will abort the transaction without completing the handshake because of a malformed message. From our assumption that $Mallory$ cannot access the DDM module, we note that properties of DDM such as uniformity give the advantage to spread the dialects evenly across all the sample requests, which makes it harder for $Mallory$ or $Eve$ to guess the successive dialect based on previous experience of dialects. The cost-based dialect selection property confers a customized neural network design to trigger a dialect with the highest frequency (which creates a vice-versa for Uniform distribution) and revamp the predictions frequently to confuse the MITM attackers.
\begin{table}[H]
\scalebox{0.8}{%
\begin{tabular}{ |p{2.5cm}|p{2.1cm}|p{2.4cm}|p{2.3cm}|}
\hline
\multicolumn{4}{|c|}{\textbf{Security analysis matrix}} \\
\hline
\textbf{Attack} & \textbf{Notable implication} & \textbf{Vulnerability} & \textbf{Requires Continuous authentication \& use of application layer features} \\
\hline
 Session hijacking attacks & MITM, packet sniffing & Rerouting the target request & $\checkmark$   \\ \hline
 Proxy attacks and firewalls (a form of MITM) & Interception software & Origin confusion issues & $\checkmark$   \\ \hline
\textcolor{black}{Context Confusion attacks} & Downgrade to lower version of TLS & Transpires because of shared TLS certificates between the servers & $\checkmark$   \\
\hline
\end{tabular}%
}
\caption{Security analysis matrix.}
\label{table1}
\end{table}
\vspace{-\baselineskip}
\section{Evaluation}
\label{eval}
In this section, we evaluate the effectiveness of Verify-Pro on three real-world protocol implementations: \textit{FTP, HTTP \& MQTT}. We analyzed numerous implementations of FTP to collect various communication transactions. We customize the FTP protocol on client-server systems to include 15 customized transactions-protocol dialects to provide continuous authentication for each request in the session. MQTT (an IOT message protocol) runs over TCP/IP. MQTT is a publish-subscribe protocol that transports messages between entities. We choose to customize the publish packet to build different customized versions of the protocol and evaluate our Verify-Pro feasibility. HTTP is an application layer protocol that runs over TCP. This protocol is similar to FTP, where the client sends the requests, and the server sends responses. The HTTP protocol is primarily used to deliver data (image files, HTML files, etc.) The default port is 80, but other ports can be used as well. HTTP is used for transferring smaller files such as web pages, whereas FTP is more efficient at transferring large files.\\
\textbf{Experiment Setup:} Our experiments are conducted on a 3.20 GHz Intel(R) Core(TM) CPU i7-4790S machine with 15.5 Gigabytes of main memory. The operating system is Ubuntu 18.04 LTS. 
\subsection{Customizing FTP}
As a target protocol for our proof-of-concept evaluation, FTP has two main benefits: (a) a light-weight network protocol having finer performance, flexibility, and ease in testing, and (b) It has less complexity in design, supports in customizing the protocol at the binary level for providing additional security measures. 
As shown in Appendix \textcolor{red}{\ref{appendix:b}}, when a request ‘$GET\ \ filename$’  is sent to the server, the default FTP protocol has a request-response handshake. After applying Verify-Pro on FTP, the PDs modules comprise various communication transactions on both sides. DDM module on both client-server systems is used to choose a dialect ‘$d_i$’ for each unique request ‘$R_i$’. Request ‘$R_i$’ (undialect-ed request) is sent to the server, and the client awaits the response from the server to verify the server identity.  On the server-side, utilizing the request ‘$R_i$’ received from the client, the server uses the DDM module to determine the dialect number ‘$d_i$’ to send a dialect-ed response ‘$resp$’ to the client. In turn, the client uses the SRV module to validate the response sent from the server- resulting in identity verification.\\
To demonstrate the effectiveness of our tool (see Appendix \textcolor{red}{\ref{spoofedornot}}), Verify-Pro, we create an attacker FTP server (A-server) that does not comprise any protocol dialects and DDM module. Once client  \& A-server systems are on the communication loop, the target request is sent from a genuine client to a malicious server. The malicious server can start fabricating and sending malicious responses to the genuine client. The malicious server fails in the dialect evolution phase, as continuous authentication is performed for every request in the session. Since the client protocol dialect pattern is dynamically changed for every request, the malicious server finds it difficult to understand the dialect evolution pattern generated by the DDM module. We believe that this method even complies with TLS encryption schemes to increase confidentiality too. We prove that our method helps defend the communication protocols from attack vectors such as rerouting the target request, using malicious proxies or interception software, and context confusion attacks. 
\vspace{-\baselineskip}
\begin{flushleft}
\begin{table}[H]
\scalebox{1.0}{
\begin{tabular}{ |p{3cm}|p{2.2cm}|p{2.2cm}|  } 
\hline
\multicolumn{3}{|c|}{\textbf{Performance analysis of Verify-Pro on FTP}} \\
\hline
Performance Index&  FTP & Verify-Pro (FTP) \\
\hline
CPU\% Utilization & $<$1\%   & 1\% \\
System time/sec & 43.871 sec  & 44.106 sec \\
DDM model time/sec &N/A (No DDM) & 0.0723 sec \\
SRV model time/sec    &N/A (No SRV) & 0.000525 sec \newline (only on client) \\
\hline
\end{tabular}}
\captionsetup{justification=centering} \caption{Performance metrics of Verify-Pro FTP \& FTP} 
\begin{flushleft}
\footnotesize $^1$System time: Time recorded from the user login to a 20 byte file transfer in seconds for Dialect 8 (Verify-Pro) \& FTP (deployed dialect-8 template). \newline $^2$DDM time: Time logged from triggering of the user request to the prediction of dialect number. \newline $^3$ SRV time:Time logged from feeding the response as input to the Decision tree until outputting the dialect as confirmation.
\end{flushleft}
\label{table:2}
\end{table}
\end{flushleft}
\vspace{-\baselineskip}
\subsubsection{\textbf{Trade-offs of DDM module.}}
\begin{figure}[H]
\centering
\includegraphics[width=9cm]{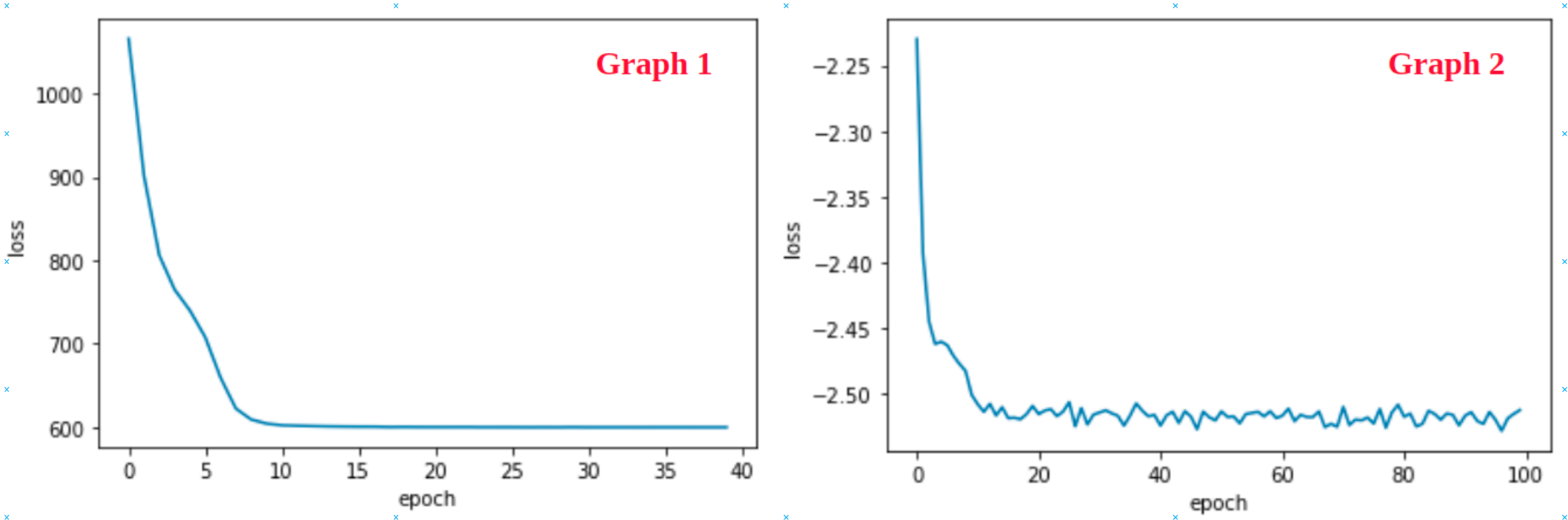}
\caption{loss vs. number of epochs.}
\small *Graph1 shows the convergence of cost property \& Graph2 shows the convergence of uniformity
\label{trade-off1}
\end{figure}

\begin{figure*}
     \centering
     \begin{subfigure}[b]{0.25\textwidth}
         \centering
         \includegraphics[width=\textwidth]{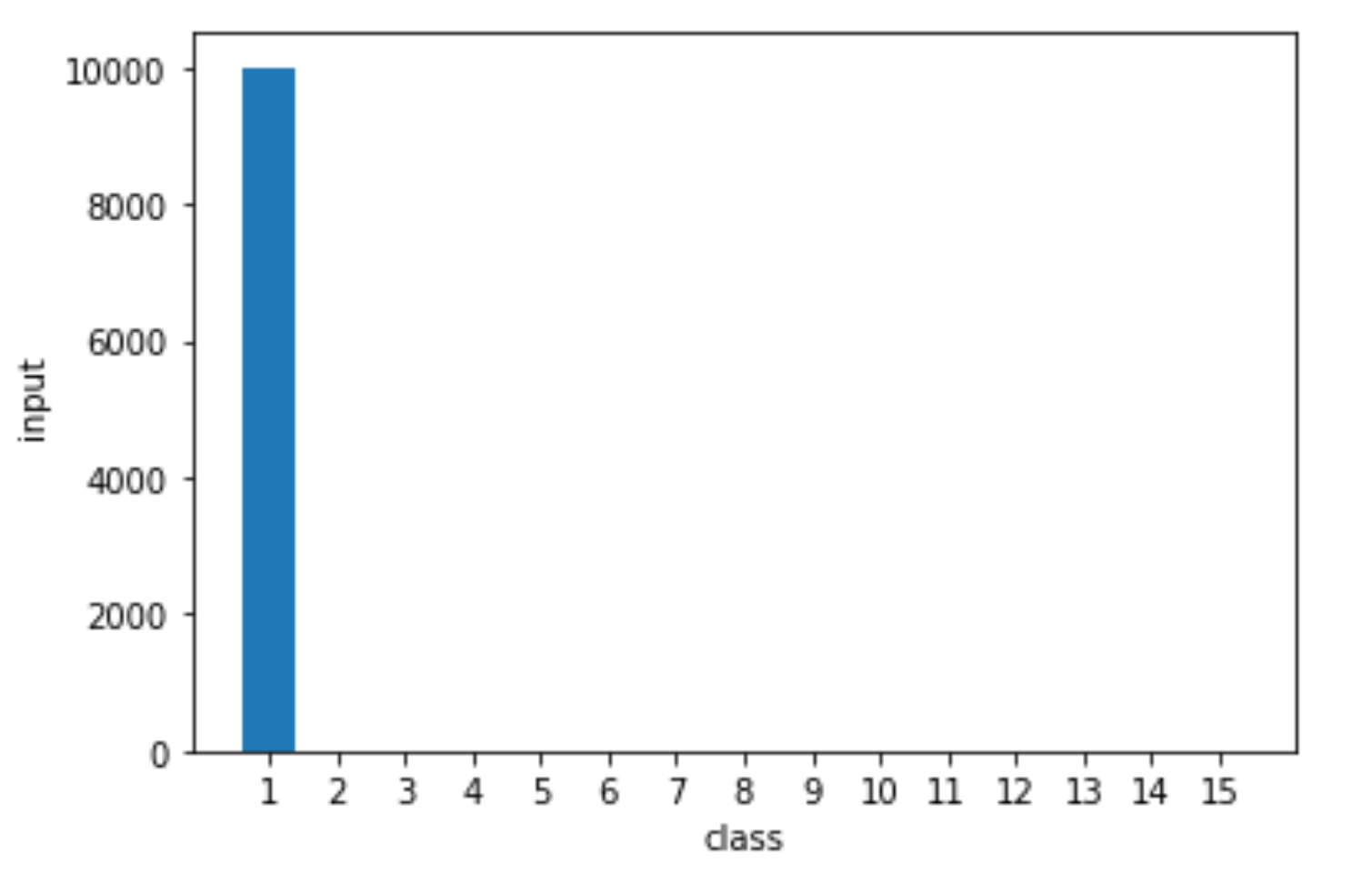}
 \captionsetup{justification=centering} \caption{\textit{a = 1 for l2 loss \newline a = 0 for l1 loss}}
         \label{cost}
     \end{subfigure}
     \begin{subfigure}[b]{0.24\textwidth}
         \centering
         \includegraphics[width=\textwidth]{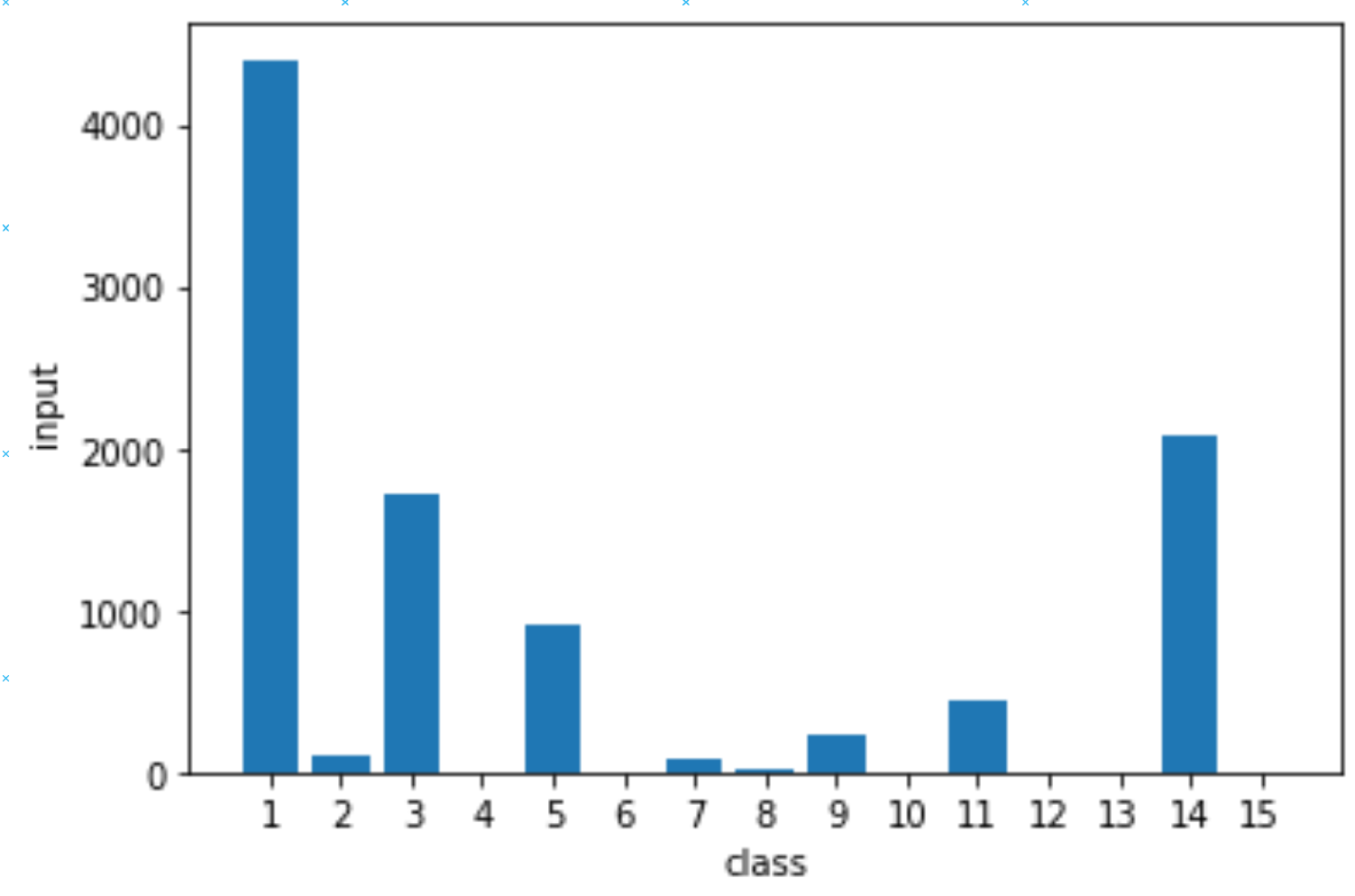}
   \captionsetup{justification=centering}      \caption{\textit{a = 0.8 for l2 loss \newline a = 0.2 for l1 loss}}
         \label{cost1}
     \end{subfigure}
     \begin{subfigure}[b]{0.25\textwidth}
         \centering
         \includegraphics[width=\textwidth]{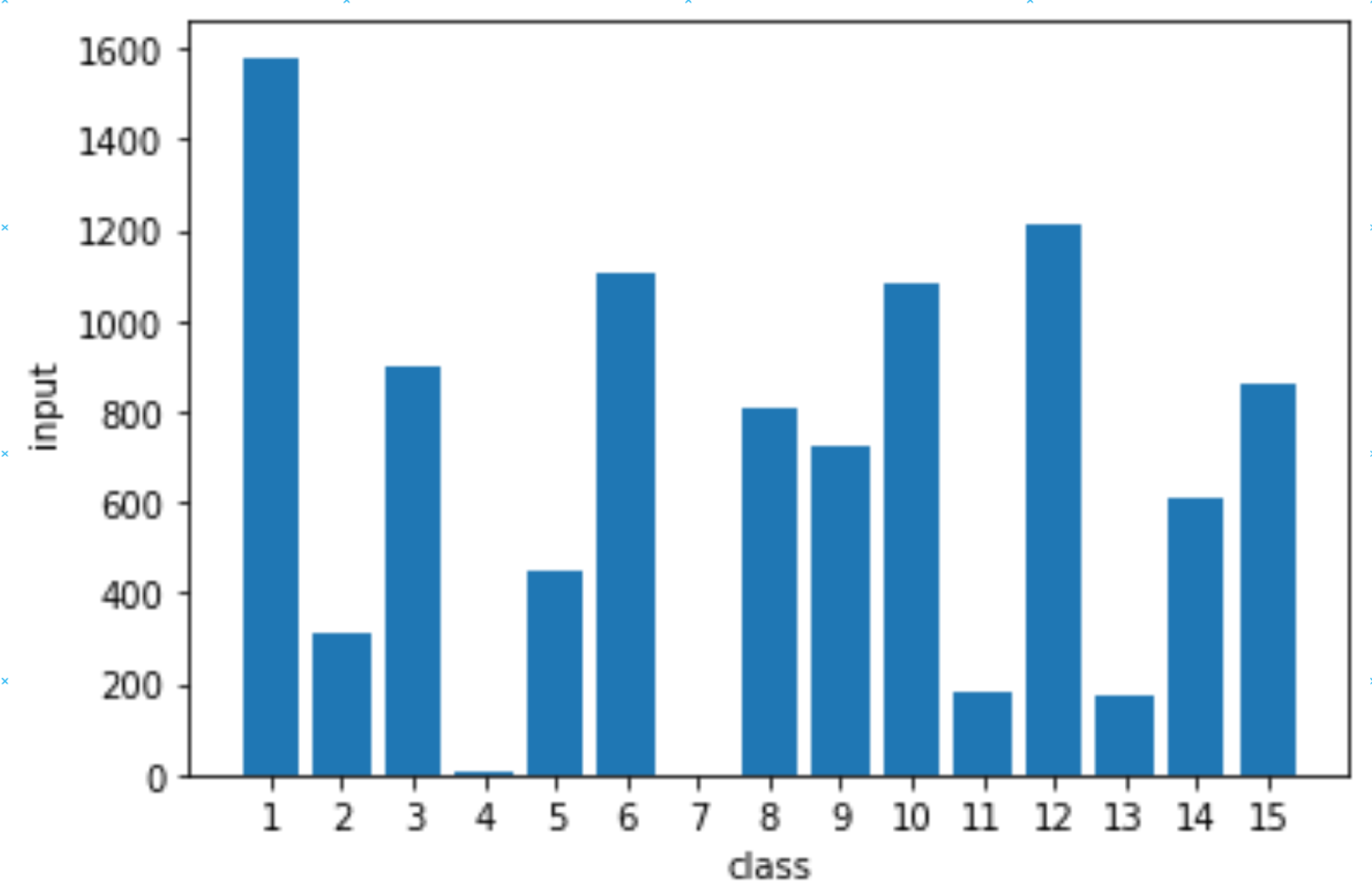}
 \captionsetup{justification=centering}        \caption{\textit{a = 0.4 for l2 loss \newline a = 0.6 for l1 loss}}
         \label{medium}
     \end{subfigure}
     \begin{subfigure}[b]{0.24\textwidth}
         \centering
         \includegraphics[width=\textwidth]{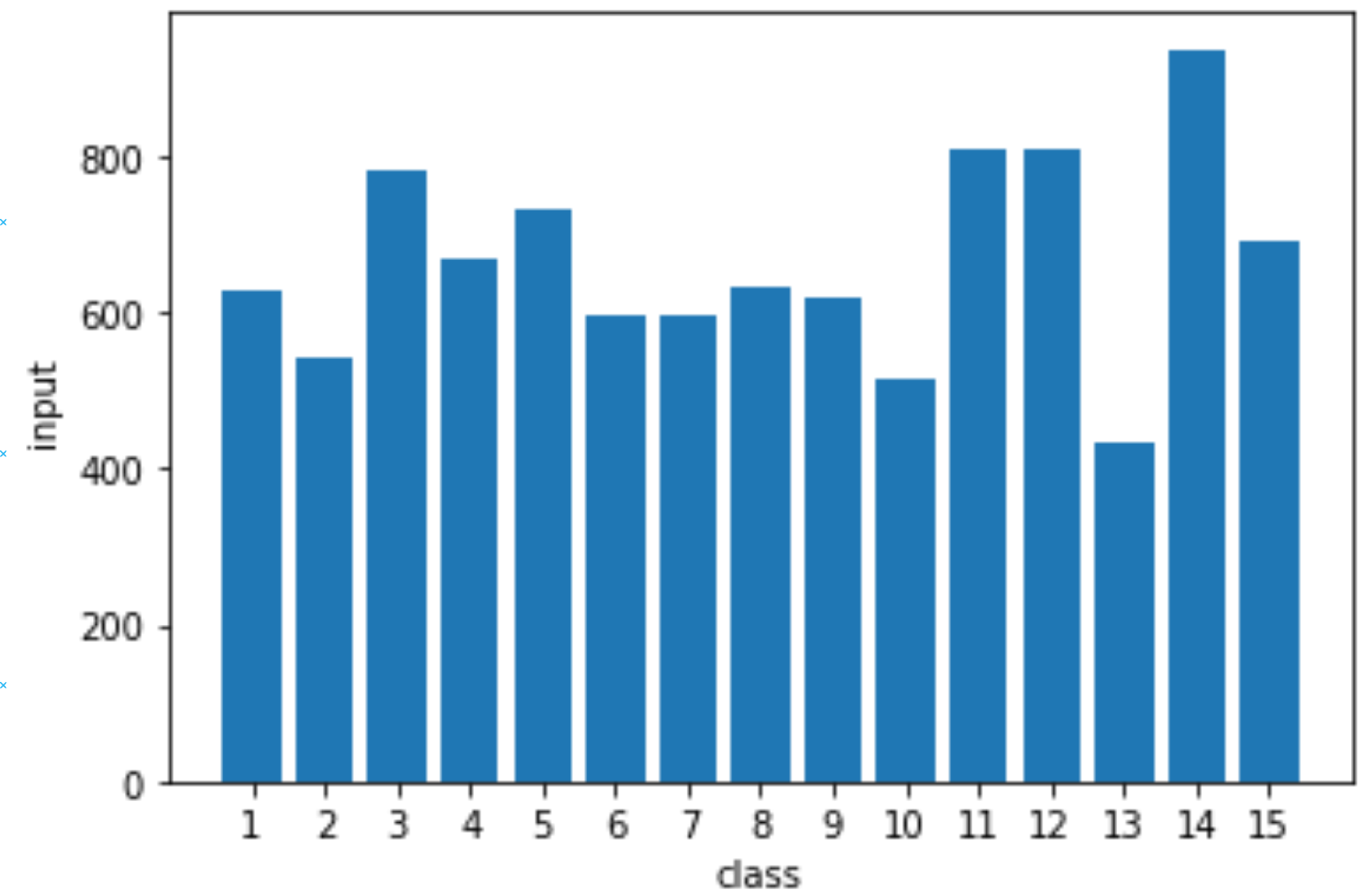}
\captionsetup{justification=centering}         \caption{\textit{a = 0 for l2 loss \newline a = 1 for l1 loss}}
         \label{uniform}
     \end{subfigure}
 \captionsetup{justification=centering}       \caption{Variations of charts considering the trade-off factor $a$. As the trade-off factor decreases the distribution starts to spread across more dialects. This graph shows the dialect numbers on $x-axis$ \& sample requests on $y-axis$.}
        \label{trade-off2}
\end{figure*}

In this section, we vary the trade-off factor $'a'$ of our deep neural network that ensures the property of flexibility and provides a trade-off between the loss l1 \& l2 properties. As can be seen from figure \textcolor{red}{\ref{trade-off1}}, graph-1 \& graph-2 demonstrate the results of convergence of cost property and uniform distribution properties. We used $128\times128\times15$-layers, 0.00001-learning rate, 40 epochs for cost loss \& 100 epochs for entropy loss, 128-batch size as the system configurations. From equation \textcolor{red}{\ref{eq3}}, we use a trade-off factor $a$ to show the flexibility \& ability of the DDM module algorithm for adapting to different requirements. For instance, consider two dialects 1 \& 7 with average sample time frames 0.0707 sec \& 8.112 sec respectively ($\S$ \textcolor{red}{\ref{timeframes}}). Our customized neural network assists in picking a dialect that has less execution overhead. Normally, we assign the cost of dialect $'C_x'$ based on the number of request-response pairs exchanged between two parties. But for experimental purposes, say, we need to pick an optimal dialect with less execution time, then we can define the cost based on our assumptions. Dialect 1 has a low cost in this experiment because of its low execution overhead compared with dialect 7. We trained using the Adam optimizer and found an optimal solution with all the samples predicting dialect $D1$ when the trade-off factor $a$ is set to $0$ for entropy loss (cost loss predicts optimal dialect as $D1$-Figure~\textcolor{red}{\ref{cost}}). Relative to this experiment, we conclude two cases: (1) We make our model more efficient by picking an optimal dialect 1, which has 11374\% less overhead than dialect 7 (based on time frames). (2) By varying the neural network weights, the architecture may trigger various dialects and confuses the middle attackers to speculate the pattern. From this, we can conclude that there exists no patterns in the sequence of dialects and this it-turn gives an added security advantage. On the other hand, we use the above-mentioned system configurations with 30 epochs to analyze how the trade-off factor $a$ affected the distribution. The interesting find from Figure \textcolor{red}{\ref{uniform}} is that, as the $a$ value decreases from $1$ to $0$, we observed that all the dialects are evenly distributed in such a way that every dialect has an equal likelihood of happening. For instance, the Figures \textcolor{red}{\ref{cost1} \& \ref{medium}} depict the variation when the trade-off factor for cost loss is changed from $0.8$ to $0.4$. Considering the cost loss with trade-off factor $a=0.8$, even though the fixed entropy loss is included, the optimal dialect $D1$ is predicted for 50\% more sample requests when compared with the second-best optimal dialect $D14$. Similarly, the cost loss with trade-off factor $a=0.4$ shows that  $D1$ still has the highest frequency, but also observed that the dialects appear to spread more evenly when compared with Figure \textcolor{red}{\ref{cost1}}. Modest problems in the learning for the consolidated loss function using a trade-off factor exists because of the factors such as network architecture, features, hyper-parameters, or training procedures. One of the prominent reasons for using a neural network is its property of reproducibility in terms of security. This ensures that given the neural network with the same architecture, the predicted dialect number ‘$D_i$’ will be the same for a given request ‘$R_i$’. The predicted dialect numbers are independent of each other, i.e.; there should not be any serial correlation between the numbers generated in a sequence. This means, given any length of preceding dialect numbers, one cannot predict the next number for a given request in the sequence by observing the given dialect numbers.\\ 
\textbf{Analysis of Table \textcolor{red}{\ref{table:2}}}. In the end, we evaluate the execution overhead of Verify-Pro, by transferring a file of 20 bytes-Dialect(8) and compare the results with standard FTP (deployed dialect-8 template). To be concrete with our evaluation, we also check the overhead of the modules which are added when compared with the original FTP implementation. We present the overhead of DDM and SRV modules. Since both these modules have pre-trained models with a size of 12MB for neural network model \& 7KB for decision tree models, the execution time of these modules is negligible. Besides the overhead of DDM, SRV modules, the remaining overhead incurs when the client verifies information of each field such as command, filename, etc. We conclude that the corresponding developments neither impact the process's slowdown nor make it easy for the attackers to obtain a shadow model. Our PDs module does not incur any overhead. The dialects are created as threads such that only one instance ‘$C_i$’ will be activated for a given dialect ‘$D_i$’ \& for the unique request ‘$R_i$’. To avoid potential statistical bias, we execute the experiment multiple times and compute the average overhead (see Table \textcolor{red}{\ref{table:2}}). Precisely, we conclude that the addition of PDs, DDM, SRV modules incurs 0.536\% overhead (from system time), which is trivial; in turn, the addition of these modules enforces continuous authentication.\\
Furthermore, the run-time overhead for all the dialects except dialect 7 is $< 1$\% (on average), which is negligible. The motive behind increasing the run-time for dialect 7 is to use it for the constraint in the dialect selection based on cost property (to depict how less run-time overhead dialects can be selected instead of the dialects which have high run-time overhead). Considering the storage overhead for the DDM module when a server is connected to multiple clients, deploying multiple neural network models occupies large amounts of disk space. To counter this problem, the advanced users can use neural network pruning \cite{manessi2018automated, han2015deep} for compressing the deep neural network and produce the same functionality. 
\vspace{-\baselineskip}
\subsection{Customizing MQTT \& HTTP}
In this sub-section, we evaluate the efficiency \& usability of communication protocols like HTTP \& MQTT after integrating with Verify-Pro.
\label{eval1}
\subsubsection{MQTT Protocol}
\begin{figure}[H]
\centering
\includegraphics[width=8.5cm]{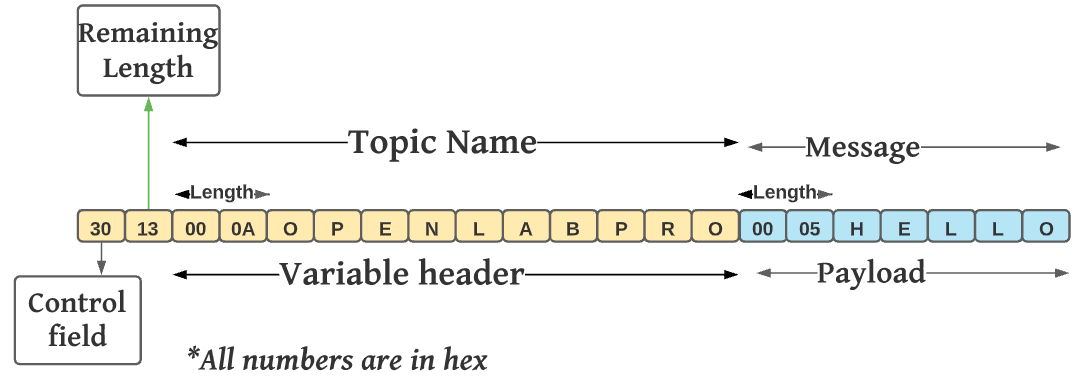}
\caption{MQTT publish packet format.}
\label{Figure4}
\end{figure}

MQTT is a standard lightweight IoT messaging protocol that is used for IoT devices. It is a binary-based protocol, which has $command$ and $command\ \ acknowledgment$ format. The MQTT protocol payload carries the data such as binary, ASCII data, etc. It uses packets of small size, hence offers benefits for low bandwidth applications. MQTT packet (Figure \textcolor{red}{\ref{Figure4}}.) contains a fixed header (including control header), variable header, and payload on the application layer.
We programmed a standard MQTT client and broker (server) and applied our Verify-Pro on them.  
The client will send a $publish$ packet to the server, and the server will respond with ‘$pub-ack$’ message. MQTT protocol dialects are presented in Table \textcolor{red}{\ref{table3}}. In this protocol, we assume the client has to prove its authenticity to the server, meaning the protocol dialect-ing actually starts from the client. As can be seen from Figure \textcolor{red}{\ref{Figure4}}, let’s consider the publish packet with the message ‘$HELLO$’ to the topic ‘$OPENLABPRO$’. In the variable header, the first 2 bytes specify the length of the topic and are then succeeded by the topic. Similarly, the payload header has the first 2 bytes denoting the length of the message, followed by the message itself. The PDs module contains dialects with mutations done to topic name, message variable headers, etc. The input of the neural network (DDM) for this protocol is the topic of the publish packet. Since the topic is used as input, the dialect number predicted from it will be used as a key to start the communication. SRV module, on the server-side, verifies whether the client is communicating in dialect ‘$D_i$’ for a topic ‘$T_i$’. 
\begin{table}[H]
\begin{tabular}{|p{4cm}|p{4cm}|}
\hline
\centering
\textbf{Dialect name} & \textbf{Mutations done}  \\ \hline
Header shuffle & Topic and message fields are switched  \\ 
\hline
Transmutation of messages & A default MQTT has 1 topic and 1 value in each publish packet, but this variant publishes the message with 2 topics and 2 values to save bandwidth\\ 
\hline
Mutation of payload & This variant divides a single packet with (1 topic \& 1 value) into two different packets  \\ 
\hline
\end{tabular}
\caption{MQTT protocol dialects.}
\label{table3}
\end{table}

To illustrate the capability of MQTT-with PDs in defending against DoS attacks, we implement an attacker client (A-client) that strives to flush the server by continuously sending ‘$publish$’ packets. The PDs module uses different variants of protocol dialects such as transmuting the handshake rules, shuffling headers, mutate the topic and value fields (mainly shows how the packet is sent). A-client fails to understand the dialect evolution based on each unique topic and the specific message formats. The A-client fails to send multiple ‘$publish$’ packets to exhaust the server’s resources, as the PDs \& DDM modules are available only with the genuine client, only a genuine client, based on its topic, can send the dialect-ed messages and the server based on the same topic can know the dialect mutation technique. The genuine client \& server can effectively communicate and publish information on the server due to the cloned DDM module, and the genuine client always knows how to send the packet in a ‘$particular\ \ dialect$’ – for unique ‘$topic$’. 
For instance, consider a topic $OPENLABPRO$ with a message $HELLO$ is triggered by a genuine client. At this point, the client already computes the dialect number $n$ using the DDM module and preserves it for future verification. The server on receiving the $publish$ packet again follows the same process by using a DDM module to compute a dialect number $n$. Dialect-ing technique, transmutation of messages mutates the topic \& messages packets by tagging dummy topic, message in a single publish packet. Then, the server prepares the dialect-ed response $resp$ and sends it to the client. Finally, the SRV module on the client side assists in verifying if the response $resp$ it received from the server has dummy topic \& message fields with original ones in the same $publish$ packet, resulting in verifying the format in which the messages are sent. 
\begin{table}[H]
\scalebox{0.9}{
\begin{tabular}{|p{2.5cm}|p{1.25cm}|p{1.25cm}|p{1.25cm}|p{1.25cm}|}
\hline
\centering
\textbf{Performance Index} & \textbf{HTTP} & \textbf{Verify-Pro(HTTP)} & \textbf{MQTT} & \textbf{Verify-Pro(MQTT)} \\ \hline
CPU Utilization &  $<$1\%   & 1\% &  $<$1\%   & $<$1\% \\ 
\hline
System time/sec  & 4.867 sec & 4.902 sec & 4.293 & 4.325\\ 
\hline
DDM model time/sec  &N/A (No DDM) & 0.0538 sec  &N/A (No DDM) & 0.0715 sec \\ 
\hline
SRV model time/sec   &N/A (No SRV) & 0.000346 sec \newline (only on client)  &N/A (No SRV) & 0.000617 sec \newline (only on server) \\ 

\hline
\end{tabular}}
\caption{Performance analysis of MQTT \& HTTP protocols}
\footnotesize $^1$System time: We compute the average time for all the dialects (HTTP \& MQTT). We only recorded the system time for one complete handshake. More precisely, the time the dialect is triggered until it gets the confirmation of dialect from another machine.
\label{tableperf}
\end{table}
\vspace{-\baselineskip}
\subsubsection{HTTP Protocol}
\label{eval2}
\begin{figure}[h]
\centering
\includegraphics[width=8.5cm]{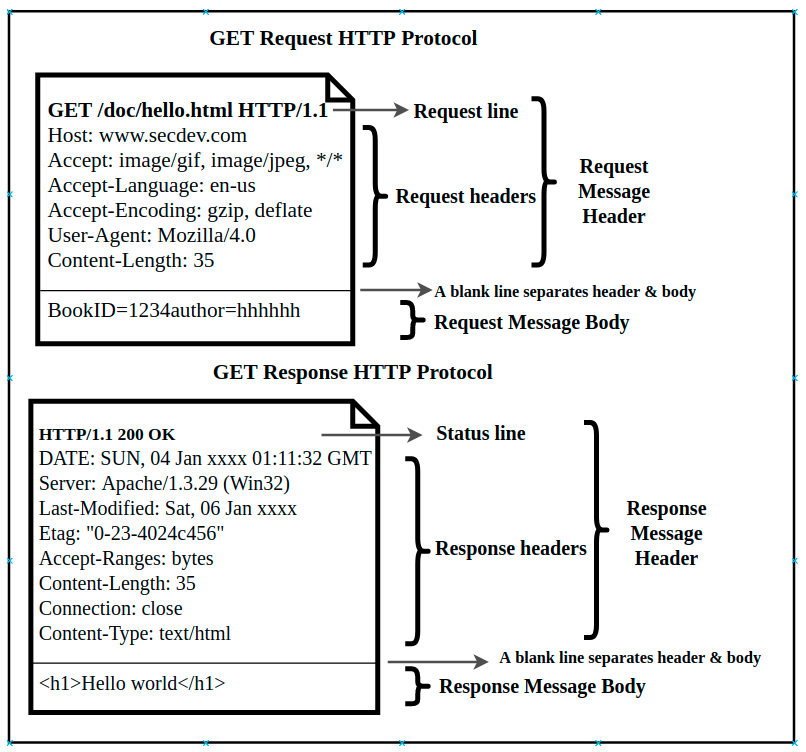}
\caption{HTTP-GET request-response packet format.}
\label{figure5}
\end{figure}

HTTP is one of the leading protocols used on the Internet. HTTP assists the users in interacting with web resources such as HTML files by transmitting messages between client and server systems. 
In this protocol, we assume the server has to prove its authenticity to the client, meaning the protocol dialect-ing starts from the server-side. Consider the scenario of a GET command, the client will send a GET request to the server, and the server will respond with a body of HTML data. As a proof of concept, to show the feasibility of Verify-Pro on HTTP, we create a couple of dialects described in Table \textcolor{red}{\ref{table4}}. As can be seen from Figure \textcolor{red}{\ref{figure5}}, the GET request is used to retrieve the HTML body as the response from the server storage. The PDs module comprises dialects with mutations performed on the message body. The input of neural network (DDM) for this protocol is the command and name of the HTML file (‘$GET\ \ hello.html$’). 
\begin{table}
\begin{tabular}{|p{4cm}|p{4cm}|}
\hline
\centering
\textbf{Dialect name} & \textbf{Mutations done}  \\ \hline
Mutation of Response message & The response message 
(HTML file in Fig.\textcolor{red}{\ref{figure5}}) is sent in 5 packets from server side  \\ 
\hline
Field shifting & For a given Request $'R_i'$, the response is sent in two packets with packet 1- two messages (HTML files) \& packet 2- no response body  \\ 
\hline
\end{tabular}
\caption{HTTP protocol dialects.}
\label{table4}
\end{table}
Table \textcolor{red}{\ref{table4}} presents the protocol dialects applied on HTTP protocol. In HTTP protocol, we focus on the response (the HTML file received) for a given request (e.g., GET hello.html HTTP/1.1). We mainly focus on preventing the MITM-session hijacking attacks (reroute the target request), Context confusion attacks, HTTP DoS attacks. Since we assume the client is ideally genuine, we tend to provide continuous authentication for each transaction such that the server verifies its identity to the client. \\ 
To demonstrate the practical feasibility of Verify-Pro, we implement an attacker HTTP server (A-server-\textit{no PDs and DDM module}). Once, the connection is established, the target request is rerouted from the HTTP client to a malicious server using MITM capabilities. The malicious server fails, as continuous authentication is performed for every request in the session. Since the client protocol dialect pattern is dynamically changed for every request, the malicious server finds it difficult to understand the dialect evolution pattern generated by the DDM module. 
SRV module is programmed in such a way that, for dialect, ‘$D_i$’ consists of pattern ‘$P_i$’. Any deviation in this pattern results in the termination of the entire session of the communication protocol. For example, consider a $GET$ request, with a $hello.html$ to be received from the server. We mutate the response message, such as sending the HTML body in five different packets. Only a genuine client knows that for a given dialect $n$, the HTML body will be sent in five packets. 
SRV module on the client-side has the required features to identify the format of the message and confirm if the message format is legitimate or spoofed. As can be seen from Table \textcolor{red}{\ref{tableperf}}, we conclude that the addition of PDs, DDM, SRV modules incurs minimal ($< 0.8\%$) overhead (from system time) on average for MQTT \& HTTP protocols for all the implemented dialects. Our solution, Verify-Pro, shows that continuous authentication can be implemented with minimal additional overhead and less cost on top of the communication protocols and can be deployed incrementally, hence making it a scalable solution. Given its security benefits, we believe that Verify-Pro can function as an additional strong protection layer in conjunction with existing authentication mechanisms.

\section{Related work}
In this section, we first discuss the existing efforts on formally analyzing the fingerprinting method to identify bots and other protocols and then compare our approach with previous works. 
Plenty of previous works are about fingerprinting.
However, most of them have different areas of focus, respectively.
Numerous studies focus on HTTPS traffic presence identification~\cite{shbair2020survey}. Hfinger~\cite{bialczak2021hfinger} a malware fingerprinting tool that extracts the information from the parts of the request such as URI, protocol information, headers, etc., and generates fingerprints. DNS protocol fingerprinting can be utilized as a process to detect DNS amplification DDoS attacks~\cite{fachkha2014fingerprinting}, identify DNS servers~\cite{kim2011effective}. 
When used for malware network traffic fingerprinting, the conferred approaches can detect spam messages, DDoS attacks and identify malware family units. Papers~\cite{hao2009detecting,venkataraman2007exploiting} concentrates on the origin of the email by exploring the features about the sender of an email (e.g., IP address) or evaluating the email is spam by approximating the geographical distance between sender and receiver.
Noteworthy papers of Stringhini et al.~\cite{stringhini2012b,stringhini2014harvester}, present a different conceptualization. Researchers focused on the analysis of network communication observed during the sending of a message by spam-bot while constructing SMTP dialects. Their results exemplify the errors in the implementation of SMTP protocol made by adversaries and hence a prospect for classification. Extension of the work, Botnet Fingerprinting~\cite{Bazydlo_2017}, the authors focus on the development of detection and classification of messages sent by spam-bots. They used SMTP dialects and made enhancements to the notion of botnet fingerprinting. Ouyang et al.~\cite{ouyang2014large} present a comprehensive analysis of spam detection through network features, like IPTTL extracted from TCP SYN packet or termination time of TCP three-way handshake. Existing methods are confined to lower-level system configurations (e.g., IP address, TCP three-way handshakes). In contrast to these works, our scheme mainly focuses on leveraging the application layer features in designing protocol dialects and enforces continuous authentication for every request in the session. Verify-Pro dynamically generates customized server replies using protocol dialects for each request, making it difficult for the attacker to launch a potential attack on a constantly self-adapting communication protocol. Moreover, we utilize a Dialect Decision Mechanism (DDM- customized neural network) to trigger a specific dialect under the premise of protocol dialects.
\vspace{-\baselineskip}
\section{Conclusion}
\label{sec:conclusion}
We presented a novel framework Verify-Pro, which aims to use the customized transactions, dialects as fingerprints during communication, and dynamically select different dialects based on a unique request in the session to compel continuous authentication. Empirical results indicate that Verify-Pro can be used to detect adversaries, assist in effective communication using protocol dialects, and improve security while incurring negligible execution overhead of 0.536\% (for FTP). While the attackers might adapt to the Verify-Pro environment by passively observing the traffic, by using reverse engineering methods, we argue that this deteriorates their performance, flexibility and incurs cost \& time complexities. We hope our research enables the community and the program developers to raise more attention to enforcing continuous authentication for the communication protocols. \\
\indent As future work, we will continue to include a variety of protocol dialects (number of dialects) and train the model to make it more robust. We will also explore automation strategies that will help us with the deployment of dialects in a scalable manner.


\bibliographystyle{IEEEtran}
\bibliography{main}

\section*{Appendix}

\section{FTP Dialects with Request-Response pairs}
\centering
\label{appendix:A}
\scalebox{0.45}{
\begin{tabular}{|l|l|}
\hline

Dialect number & Request-Response pairs  \\ \hline
    \    1 & Request: get filename\\
        &  Request: Port number\\
        &  \textcolor{red}{Response: File exists, file size }\\  
        &  \textcolor{red}{File sharing - Only after client verifies server's response}\\  \hline
    \    2 & Request: get filename\\
        &  Request: Port number\\
        &  \textcolor{red}{Response: file size, file size \Comment{file size sent two times} }\\ 
       &  \textcolor{red}{Response: Connection Closed}\\ \hline
        \ 3 & Request: get filename\\
        &  Request: Port number\\
        &  \textcolor{red}{Response: File exists, file size, file name}\\  \hline
    \ 4 & Request: get filename\\
        &  Request: Port number\\
        &  \textcolor{red}{Response: File size/2}\\
        &  \textcolor{red}{Response: Remaining file size}\\  \hline
    \ 5 & Request: get filename\\
        &  Request: Port number\\
        &  \textcolor{red}{Response: 1 (file exists), length of filename,
                        length of command} \\
        &  \textcolor{red}{Response: File size}\\  \hline
   \ 6 & Request: get filename\\
        &  Request: Port number\\
        &  \textcolor{red}{Response: File exists} \\
        &  \textcolor{red}{Response: length of filename,
                        length of command} \\
        &  \textcolor{red}{Response: File size}\\  \hline
    \ 7 & Request: get filename\\
        &  Request: Port number\\
        &  \textcolor{red}{Response: File exists}\\
        &  \textcolor{red}{Response: Length of filename}\\
        &  \textcolor{red}{Response: Length of command} \\  \hline
    \ 8 & Request: get filename\\
        &  Request: Port number\\
        &  \textcolor{red}{Response: File exists, size of file, filename,command}\\  \hline


\ 9 & Request: get filename\\
        &  Request: Port number\\
        &  \textcolor{red}{Response: File exists, size of file}\\
        &  \textcolor{red}{Response: filename, command} \\  \hline
\ 10 & Request: get filename\\
        &  Request: Port number\\
        &  \textcolor{red}{Response: File exists} \\
        &  \textcolor{red}{Response: size of file,filename,command} \\  \hline
\ 11 & Request: get filename\\
        &  Request: Port number\\
        &  \textcolor{red}{Response: File exists} \\
        &  \textcolor{red}{Response: size of file} \\
        &  \textcolor{red}{Response: Filename, length of filename} \\
        &  \textcolor{red}{Response: command, length of command} \\  \hline
\ 12 & Request: get filename\\
        &  Request: Port number\\
        &  \textcolor{red}{Response: size of file} \\ \hline 
\ 13 & Request: get filename\\
        &  Request: Port number\\
        &  \textcolor{red}{Response: File exists, filename,command} \\
        &  \textcolor{red}{Response: length of filename} \\  \hline
\ 14 & Request: get filename\\
        &  Request: Port number\\
        &  \textcolor{red}{Response: File does not exist}\\
        &  \textcolor{red}{Response: -(size of file)} \\  \hline
\ 15 & Request: get filename\\
        &  Request: Port number\\
        &  \textcolor{red}{Response: File exists} \\
        &  \textcolor{red}{Response: size of file} \\
        &  \textcolor{red}{Response: Filename} \\
        &  \textcolor{red}{Response: command} \\
        &  \textcolor{red}{Response: Length of command} \\  \hline
        
\hline
\end{tabular}
}
\\

\leftskip0cm\relax
\rightskip0cm\relax
\textbf{FTP Dialects.} We create a variety of dialects~Appx.~\textcolor{red}{\ref{appendix:A}} using techniques by changing communication rules such as packet mutation, generate different request-responses, communicating in numbers only. For example, in dialect 5, the communication happens with numerals - 1 (file exists), inverse communication (explained in dialect 14, where the file is transferred, but the protocol handshake is assumed to be inverse of the process). Dialect 7 includes statistics such as throughput time \& divide the file and send in sub-packets to intentionally increase the handshake time. Our main aim is to detect the adversary in the initial phase of a handshake so that the client can terminate the connection without even entering the file transferring process. As this work is a proof of concept, we only deployed 15 dialects, but when the dialects number increases, the adversary finds it even more difficult to guess the correct dialect. Table~\ref{table6} represents the time taken for each dialect to complete the handshake on client and server machines. Modest variations exist because of a variety of data transfer variations. Apart from the server's response, we also create variations in the client's response after the verification. Further, the server sends the file in different ways (divide the file into multiple data blocks, compression, etc.) will be explored in future work to prove two-way authentication.  
\vspace{-\baselineskip}
\vspace{-\baselineskip}
\subsection{\textbf{Dataset Description for DDM \& SRV modules}}
\label{appendix:a.1}
\vspace{-\baselineskip}
\leftskip0cm\relax
\rightskip0cm\relax
\subsubsection{Dataset Description for DDM} We make use of the Natural language corpus of words (https://norvig.com/ngrams/) for creating a customized dataset. Our neural network requires the input to be the $"request"$ of the communication protocol. We constructed a new candidate training set by using a list of real-world words in the NLP corpus. For instance, consider an FTP protocol. Our dataset should have a “get filename” format, we added a command – get and file extension to the words. We use this customized dataset as the main training and evaluation dataset. Doing so allows us to examine the generalization properties of the model for novel unseen examples. The dataset only includes the list of requests (e.g., “get filename”) as the model is constructed unsupervised setting, as it contains a large set of 150K unlabeled sample requests. In this work, we tend to design a customized neural network without ground truth. Based on the properties we use in the loss functions (uniformity, cost as a criterion for dialect selection, and consolidated loss), the model will have to predict the dialect numbers for each request. \\
\subsubsection{Dataset Description for SRV} Our goal is to train a classifier using this type of data and effectively construct a decision tree that will classify the samples. We only need the pattern and data type of each field of the response, and we used a python script to generate the dataset with 150,000 samples. We created a standard data set using a python script (of 150 lines of code) to generate random strings and integers in that particular packets and fields. Each data has 31 features, and the total number of labels is 15. In the SRV verification mechanism, the client only checks the pattern (structure of response) sent by the server (we create different responses for every dialect in the FTP protocol to avoid overlapping dialects).
\section{Original FTP \& Modified FTP Conversations}
\label{appendix:b}
\fbox{\begin{minipage}{25em}
\textbf{Default:}\\
Client: get hello.txt  \Comment{Download hello.txt from server disk storage}\\
Server: 226 Transfer completed \Comment{File is transferred}\\
\\
\textbf{Modified: Example of Dialect 10}\\
Client: get hello.txt  \Comment{Download hello.txt from server disk storage}\\
Server: File exists \\
Server: 20,hello.txt,get \Comment{20 is file size in bytes, hello.txt is filename, get is the command}\\
Client: Ready to receive the file.\\
Client: Connection closed, file received.\\
\end{minipage}}

\subsection{Legitimate FTP client \& Legitimate server vs. Legitimate FTP client \& Malicious server (A-server) communication templates}
\label{spoofedornot}
\fbox{\begin{minipage}{25em}
\textbf{Client:} 
ftp$>$ get hello.txt \\
    using dialect 9  \\     
    (Control) Got connection from ('127.0.0.1', 60068) \\
    200 PORT command successful \\ 
    Time taken: 1.056454 sec\\  
    File received successfully\\
\textbf{Server:}\\ 
request: ['rget','hello.txt'], dialect:9\\
using dialect D9\\
(Control) Data port is 33345\\
Sending file hello.txt to client 127.0.0.1\\
Time taken: 1.00464363 sec\\
\noindent\rule[0.5ex]{\linewidth}{1pt}
\textbf{Client:} 
ftp$>$ get hello.txt \\
    using dialect 12  \\     
    (Control) Got connection from ('127.0.0.1', 56768) \\
    200 PORT command successful \\ 
    Dialect mismatch, server didn't communicate in correct dialect\\
    Time taken: 0.00056454 sec\\       
\textbf{\textcolor{red}{A-server:}}\\ 
request:['rget','hello.txt']\\
sending a malformed message $\rightarrow$ \textit{File does not exist} \Comment{No dialects and DDM module. So the malicious server does not know what is the message format for dialect 9 \& there is no DDM module to which dialect should be used for each request} \\
\Comment{Unsuccessful transaction}
\end{minipage}}
\begin{flushleft}
\justify
*These conversations show the templates of genuine client-server and malformed server communication transactions. For example, the conversation at the top indicates a ‘$successful$’ transaction with the time (taken by client and server) displayed, dialect usage, hostname, and port numbers. In contrast, the bottom part indicates the failed transaction when communicating with an attacker server. 
\end{flushleft}

\begin{table*}
\scalebox{1.0}{
\begin{tabular}{ |p{3.2cm}|p{1.5cm}|p{1.2cm}|p{2cm}|p{3.1cm}|p{3cm}|p{1.5cm}|}
\hline
\multicolumn{7}{|c|}{\textbf{Comparison of existing counter measures with Verify-Pro}} \\
\hline
\textbf{Countermeasures} & \textbf{Difficulty} & \textbf{Cost} & \textbf{Continuous authentication} & \textbf{Features} & \textbf{Compatibility with multiple protocols} & \textbf{Automation}\\
\hline\hline
 Spam detection \cite{beverly2008exploiting,kakavelakis2011auto} & Medium & Medium & \textcolor{red}{$\Large \times$} & Transport level features& No (only for SMTP) & Yes \\ 
 SNARE\cite{hao2009detecting} & High & High &  \textcolor{red}{$\Large \times$} & Network level features& No & Yes \\
 GTID\cite{radhakrishnan2014gtid} & High & High & \textcolor{red}{$\Large \times$}& IP-level streams& Yes & Yes\\
 Botlab\cite{john2009studying} & Medium & Medium & \textcolor{red}{$\Large \times$} & Network level & No & Yes\\
Verify-Pro[\textit{this paper}] & Medium & Low & \textcolor{red}{$\Large \checkmark$} & Application layer features & Yes & Yes\\ 
 \hline
\end{tabular}}
\caption{Comparison of existing counter measures with Verify-Pro.}
\label{table5}
\end{table*}
\section{Comprehensive details}
\label{appendix:c}

\subsection{Analysis of Verify-Pro with existing techniques}
In this section, Table \textcolor{red}{\ref{table5}} describes the existing countermeasures against MITM-session hijacking, proxies/interception software, spammers etc, regarding the deployment considerations. Difficulty refers to the overall evaluation of integrating the corresponding PDs, DDM, SRV modules, and monetary, man-power, and time costs. The compatibility feature indicates the number of communication protocols compatible with the technologies proposed.

\subsection{Concrete time frames of dialects in FTP}
\label{timeframes}
\begin{flushleft}
\begin{table}[H]
\begin{tabular}{ |p{2cm}|p{5cm}| }
\hline
\multicolumn{2}{|c|}{\textbf{Time frames of dialect templates}} \\
\hline
Dialect number & Whole handshake time (sec) \\
\hline
Dialect 1 & Client:0.0623 sec \& Server:0.00841 sec   \\
Dialect 4   & Client:0.0931 sec \& Server:0.0083 sec \\
Dialect 7 & Client:4.085 sec \& Server:4.0556 sec  \\
Dialect 8 & Client:0.0075 sec \& Server:0.0017 sec\\
Dialect 9 & Client:0.0815 sec \& Server:0.00868 sec  \\
Dialect 10 & Client:0.017 sec \& Server:0.049 sec  \\
Dialect 12 & Client:0.052 sec \& Server:0.00856 sec\\
\hline
\end{tabular}
\caption{Various dialects with respective time frames.}
\label{table6}
\end{table}
\end{flushleft}
\vspace{-\baselineskip}
\vspace{-\baselineskip}
\vspace{-\baselineskip}
We enumerate the time frames, for particular dialects-from calling the function \textit{do\_get} (get command) on client, calling the function \textit{send\_file} on server, until the file transfer of 20 bytes to the client, for each dialect-\textit{time} library from python package is imported to note the accurate time to show the difference between some dialects of Verify-Pro in Table \textcolor{red}{\ref{table6}}. From the table, we infer that the dialects (Verify-Pro) with their mutations vary in the execution time (negligible execution overhead difference in the execution path for each dialect template). In the view of Verify-Pro, only one dialect $D_i$ can be executed for unique request $R_i$. Hence, there won't be any background execution overhead that might impact the already processing dialect. We refer to the FTP protocol: \textit{https://github.com/ShripadMhetre/FTP-Python} as the standard FTP. We execute the commands with the same file size 20 times and compute the average to avoid potential bias. We conclude that our additional developments (with SRV module, different request-responses) do not deteriorate the standard FTP by incurring execution overhead, as all the additional PDs are spawned as separate instances.

\setlength{\arrayrulewidth}{0.3mm}
\setlength{\tabcolsep}{1.0pt}
\renewcommand{\arraystretch}{1.5}

\end{document}